\documentclass[twocolumn,showpacs,preprintnumbers,amsmath,amssymb]{revtex4}

\usepackage{graphicx}
\usepackage{dcolumn}
\usepackage{bm}
\usepackage{braket}
\usepackage{upgreek}

\begin{document}


\title{Derivative Expansion of Wave Function Equivalent Potentials}

\author{Takuya Sugiura$^1$}
 \email{sugiura@rcnp.osaka-u.ac.jp}
\author{Noriyoshi Ishii$^1$}
\author{Makoto Oka$^{2,3}$}
\affiliation{%
  $^1$Research Center for Nuclear Physics (RCNP), Osaka University,
      Osaka 567-0047, JAPAN \\
  $^2$Department of Physics, Tokyo Institute of Technology,
      Tokyo 152-8551, Japan \\ 
  $^3$Advanced Science Research Center, Japan Atomic Energy Agency,
      Tokai, Ibaraki, 319-1195, Japan
}%

\date{\today}

\begin{abstract}
Properties of  the wave  function equivalent potentials  introduced by
HAL   QCD   collaboration   are   studied   in    a    non-relativistic 
coupled-channel model.
The derivative expansion is generalized, and then applied to the
energy-independent and non-local potentials. The expansion
coefficients are determined from analytic solutions to the
Nambu-Bethe-Salpeter wave functions.
The scattering phase shifts computed from these potentials are compared
with the exact values to examine the convergence of the expansion.
It is confirmed that the generalized derivative expansion converges in
terms of the scattering phase shift rather than the functional
structure of the non-local potentials.
It is also found that the convergence can be improved by tuning either
the choice of interpolating fields or expansion scale in the
generalized derivative expansion.
\end{abstract}



\maketitle

\section{\label{sec:INTRO}INTRODUCTION}

The  nuclear  force, or  the  nucleon-nucleon  (NN) potential,  is  of
crucial  importance in  nuclear physics.   It serves  as an  essential
building block  to understand  the structure  and reactions  of atomic
nuclei.
It is also used to  study the  equation of states  of the  nuclear matter,
which  provides important  information  about  supernova explosions  and
neutron star structure.
Today, several types of phenomenological nuclear forces have been
provided, which precisely describe a large number of experimental data
of NN scattering and deuteron properties
\cite{Machleidt:2000ge,Stoks:1994wp,Wiringa:1994wb}.
Chiral effective field theory has also made significant progress in
determining the nuclear force \cite{Epelbaum:2008ga}.

Theoretical derivation of the nuclear force has long been a challenge.
The problem is that quantum chromodynamics (QCD), the ultimate theory
of the strong interaction, shows a number of non-perturbative aspects
at low energies.
Lattice QCD Monte Carlo calculation provides a promising framework to
study the non-perturbative phenomena of QCD, such as hadron-hadron
scattering.
The standard methods to study the scattering phenomena is
L\"uscher's finite volume method \cite{Luscher:1990ux}, which has been
extensively applied to the NN system
\cite{Fukugita:1994ve,
Beane:2006mx,Beane:2011iw,Beane:2012vq,Orginos:2015aya, 
Yamazaki:2011nd,Yamazaki:2012hi,Yamazaki:2015asa, 
Berkowitz:2015eaa,
Iritani:2016jie}.

Recently, a method has been proposed by HAL QCD collaboration
\cite{Ishii:2006ec,Aoki:2009ji,HALQCD:2012aa} to determine the nuclear
force based on lattice QCD.
It has been applied to many targets,
including nucleon-hyperon (NY), YY, and NNN interactions~\cite{Aoki:2012tk}.
In the HAL QCD method, the Nambu-Bethe-Salpeter (NBS) wave functions
play a central role. The nuclear force is defined as an
energy-independent {\em ``wave function equivalent potential''} by
demanding that the Schr\"odinger equation should reproduce the NBS
wave functions for many different energy levels.
The asymptotic long-distance behavior of the NBS wave functions
ensures that the potential thus obtained reproduces the scattering
phase shift correctly \cite{Aoki:2009ji}.
Indeed, it is shown in Ref.~\cite{Kurth:2013tua} that the phase shifts
from the HAL QCD method agree quite well with those from L\"uscher's
finite volume method in the $\pi\pi (I=2)$ system.
Although different choice of interpolating fields leads to different
NBS wave functions and accordingly different potentials, all of them
lead to unique scattering phase shift.

In many cases, a potential is assumed to be a local single multiplication
operator $V(\vec r)$, so that it is expressed as $\langle \vec r | V |
\vec r' \rangle =V(\vec r)\delta(\vec r - \vec r')$ in the coordinate
space.
In general, however, it is not possible to demand that such an
energy-independent and local operator reproduce a set of NBS wave
functions for various different energies simultaneously.
Thus, in the HAL QCD method, the potential is considered to be
non-local as an integration operator.
The proof is given in Ref.~\cite{Aoki:2009ji}, where it is shown that
such an energy-independent and non-local potential actually exists.
The non-locality of a HAL QCD potential is taken into account by the
derivative expansion: the potential is expressed as a power series of
spatial derivatives, coefficients of which are energy-independent and
local functions.
For instance, the leading order of the nuclear force consists of the
central and tensor forces, and the next-leading order of the
spin-orbit forces.

In practice, potentials with higher order derivatives are often
inconvenient for either application or their numerical construction.
It is thus desirable to improve the convergence of the derivative
expansion.
A possible strategy is to suitably tune the interpolating
fields. However, it is not straightforward to follow this strategy for
the following reasons.
First, calculation  with  varying  interpolating  fields
requires additional numerical costs.
Secondly, brute force evaluation of the convergence is challenging;
while NBS wave functions at several energies with sufficient accuracy
are necessary in order to implement the derivative expansion, only a
limited number of excited states are accessible in lattice QCD, with
increasing uncertainty for higher excited states.

In order to make an alternative evaluation of the convergence, a
previous study examined the ``energy dependence'' of the HAL QCD potential
\cite{Murano:2011nz}.
They employed the standard local interpolating fields for the nucleon
and computed leading order HAL QCD potentials from the NBS wave
functions for $E\sim 0$ MeV and $E\sim 45$ MeV, where the two cases
were realized as the ground states under the periodic and the
anti-periodic spatial boundary conditions, respectively.
Since the derivative expansion was truncated at the lowest order, the
energy independence of the potential was only approximate.
Results showed that the discrepancy between the two potentials is
negligibly small, indicating that these energy-dependent local
potentials can be regarded as energy-independent local in this energy
region.
In  this way,  as far  as  the NN  potential is  concerned, the  local
standard interpolating  field turned  out to  lead to a potential with
small non-locality in the low-energy region.

Even so, non-locality of a HAL QCD potential may play an important
role in describing hadron-hadron scattering in higher energy region.
It is desirable to establish general methodology to evaluate the
non-locality and to improve the convergence of the derivative expansion.
In this paper, we investigate the properties of HAL QCD potentials
when the derivative expansion is explicitly performed to higher
orders.
Since its numerical evaluation requires precision study, we employ a
1+1 dimensional non-relativistic coupled-channel model introduced by
M.~Birse \cite{Birse:2012ph}, which provides analytic NBS wave functions.
We generalize the derivative expansion to avoid a trouble in applying
the na\"ive expansion to the model with unsmooth NBS wave functions.
The generalized derivative expansion  has several favorable features,
which  can  conveniently be  used  in  lattice QCD  Monte  Carlo
calculations as well.
The convergence of the generalized derivative expansion is discussed
from the viewpoint of potential structure and of scattering phase
shift.
The possibility of improving the convergence by tuning the choice of
interpolating fields is explicitly examined.

The paper is organized in the following way.
A  coupled-channel  model  of  two-body scattering  is  introduced  in
section  \ref{sec:BIRSE}.   The model  allows for simulating the
variation of interpolating fields in a particular way.
In  section  \ref{sec:POTENTIAL},  the   HAL  QCD  method  is  briefly
reviewed, followed  by generalization  of the derivative  expansion.
The results of our numerical calculation is presented in section
\ref{sec:RESULTS}. The convergence of the generalized derivative expansion
is examined, and we discuss improving the convergence.
Finally, we give our conclusions and outlook in section
\ref{sec:CONCLUSION}.

\section{\label{sec:BIRSE}Birse MODEL}

We consider a non-relativistic system in 1+1 dimensional space-time,
described by the following second-quantized Hamiltonian:
\begin{align}
\begin{split}
  \hat{H} =& \hat{H}_p + \hat{H}_n + \hat{H}_{n'} 
  + \hat{V}_{nppn} + \hat{V}_{nppn'} + \hat{V}^\dagger_{nppn'},
  \label{eq:2nd-quantized-hamiltonian}
  \\
  \hat{H}_p 
  &= \int dx\, \hat{p}^\dagger(x)\left(-\frac{1}{2M}\frac{d^2}{dx^2}\right)\hat{p}(x),\\
  \hat{H}_n 
  &= \int dx\, \hat{n}^\dagger(x)\left(-\frac{1}{2M}\frac{d^2}{dx^2}\right)\hat{n}(x),\\
  \hat{H}_{n'} 
  &= \int dx\, \hat{n}'^\dagger(x) \left(\Delta-\frac{1}{2M}\frac{d^2}{dx^2}\right) 
  \hat{n}'(x), \\
  \hat{V}_{nppn}
  &= \iint dxdy\, \hat{n}^\dagger(x)\hat{p}^\dagger(y)V_0(x-y)\hat{p}(y)\hat{n}(x), \\
  \hat{V}_{nppn'} 
  &= 2g \int dx\, \hat{n}^\dagger(x)\hat{p}^\dagger(x)\hat{p}(x)\hat{n}'(x),
\end{split}
\end{align}
where $\hat p(x)$,  $\hat n(x)$, and $\hat n'(x)$  denote scalar boson
fields.
They are analogous to the fields of proton, neutron, and an excited
neutron with excitation energy $\Delta$, respectively.  For
simplicity, we do not consider proton excitations or higher excited
states of neutron.
For Galilei covariance, p, n and n$'$ are assumed to have the same
non-relativistic mass $M$.
These fields should satisfy equal-time commutation relations
\begin{equation}
\left[ \hat{p}(x),\hat{p}^\dagger(y) \right] =
\left[ \hat{n}(x),\hat{n}^\dagger(y) \right] =
\left[ \hat{n}'(x),\hat{n}'^\dagger(y) \right] = \delta(x-y).
\end{equation}
All the other combinations vanish.
The pn-pn interaction $\hat{V}_{nppn}$ is given by a square-well
potential:
\begin{align}
V_0(x) = \begin{cases} 
-V_0 & \text{for}\hspace{3mm} |x|<R  \\
0    & \text{for}\hspace{3mm} |x|>R. \end{cases}
\label{eq:square-well}
\end{align}
Note  that  $\hat  H$  is  Hermitian  and  has  the  translational
invariance,  the  spatial  reflection invariance,  the  time  reversal
invariance, and Galilei covariance.

The non-relativistic vacuum is defined by
\begin{equation}
  \hat p(x)|0\rangle = \hat n(x)|0\rangle = \hat n'(x)|0\rangle = 0,
\end{equation}
so that it satisfies
\begin{equation}
  \hat{H}\ket{0}=0.
  \label{eq:vacuum}
\end{equation}
We consider the two-particle energy eigenstate $\ket{\Psi}$ of the
pn-pn$'$ coupling system in the center of mass frame with eigenvalue
$E$:
\begin{equation}
  \hat{H}\ket{\Psi}=E\ket{\Psi}.
  \label{eq:2particlestate}
\end{equation}
The two-particle state is parameterized by using wave functions
$\psi_0(x)$ and $\psi_1(x)$ according to
\begin{align}
\begin{split}
  \ket{\Psi} =
  & \int dx\,dy\, 
    \left\{ 
    \hat{p}^\dagger(x)\hat{n}^\dagger(y)  \ket{0} \psi_0(x-y) \right. \\
  & + \left.
    \hat{p}^\dagger(x)\hat{n}'^\dagger(y) \ket{0} \psi_1(x-y)
    \right\}.
\end{split}
\end{align}
Conversely, the wave functions can be expressed by matrix elements
\begin{align}
\label{eq:psi0psi1}
\begin{split}
\psi_0(x) &\equiv \Braket{0 | \hat{p}(x+y)\hat{n}(y)  | \Psi}, \\
\psi_1(x) &\equiv \Braket{0 | \hat{p}(x+y)\hat{n}'(y) | \Psi}.
\end{split}
\end{align}

The wave functions satisfy the following equations:
\begin{align}
\begin{split}
\left[-\frac{1}{M}\frac{d^2}{dx^2}+V_0(x)-E\right]\psi_0(x) + 2g\delta(x)\psi_1(x)
&= 0, \\
\left[-\frac{1}{M}\frac{d^2}{dx^2}+\Delta-E\right]\psi_1(x) + 2g\delta(x)\psi_0(x)
&= 0.
\end{split}
\label{eq:CCeq}
\end{align}
The derivation follows from sandwiching
$[\hat{p}(x+y)\hat{n}(y),\hat{H}]$ and
$[\hat{p}(x+y)\hat{n}(y),\hat{H}]$ with $\Bra{0}$ and $\ket{\Psi}$.
Equations~\eqref{eq:CCeq}  are identical  to the  coupled-channel equations
introduced in Ref.~\cite{Birse:2012ph}.
They can be solved analytically, which enable us to implement the
derivative expansion to higher orders with precision.
Throughout this paper, we focus on the elastic energy region
$E<\Delta$, so that the $\psi_1$ channel is closed:
\begin{equation}
  \psi_1(x) \sim e^{-\gamma |x|} , \label{eq:closed_suppressed}
\end{equation}
where $\gamma \equiv \sqrt{M(\Delta - E)}$.

A general neutron interpolating field couples to both n and n$'$. 
Let $\hat{\phi}_q(x)$ be such a field given as
\begin{equation}
  \hat{\phi}_q(x) \equiv \hat{n}(x)+q\hat{n'}(x),
  \label{eq:BirseBSwave}
\end{equation}
where $q$ is a real parameter introduced to arrange the mixing.  We
will refer to $q$ as the {\em field admixture parameter} hereafter.
The NBS wave function for p$\upphi$ is given by a linear combination
of the wave functions~\eqref{eq:psi0psi1}:
\begin{align}
\begin{split}
  \Psi_q(x)
  &\equiv
  \Braket{0|\hat{p}(x+y)\hat{\phi}_q(y)|\Psi}
  \\
  &=
  \psi_0(x) + q \psi_1(x).
\end{split}
\label{eq:NBS-Birse}
\end{align}
In the following, we use $\Psi_q(x)$ as an input of the Schr\"odinger
equation to construct HAL QCD potentials.  The interpolating field
dependence can be studied by varying the field admixture parameter
$q$.

Note that the asymptotic behavior of $\Psi_q(x)$ is independent of $q$
because $\psi_1(x)$ vanishes at large distances as
in Eq.~\eqref{eq:closed_suppressed}:
\begin{equation}
  \Psi_q(x) \underset{|x|\to\infty}{\longrightarrow} \psi_0(x)
  \simeq A \cos(k|x| + \delta(k)),
  \label{eq:as-behavior}
\end{equation}
where   $\delta(k)$   denotes   the   scattering   phase   shift   and
$k\equiv\sqrt{ME}$ is the asymptotic momentum.
As we restrict  ourselves to the parity-even sector  in this paper,
$\delta(k)$ is defined as the deviation from $\cos(k|x|)$.
The relation~\eqref{eq:as-behavior} is analogous to the lattice QCD
case, where similar asymptotic behavior is derived by utilizing the
LSZ reduction formula \cite{Aoki:2009ji,Aoki:2005uf,Lin:2001ek}.
The asymptotic behavior ensures that  the HAL  QCD potentials  are
faithful to the scattering phase  shift independently of the choice of
interpolating fields.

\section{\label{sec:POTENTIAL}Formalism}

\subsection{\label{sec:HAL}HAL QCD Potentials and Derivative Expansion}

In this section, we construct the HAL QCD potential that describes the
pn scattering problem in the elastic energy region $E<\Delta$.
To start with, consider the stationary Schr\"odinger equation in a finite box:
\begin{equation}
\left(-H_0+E_m\right)\Psi(x;E_m) = \int dx'\,V(x,x')\Psi(x';E_m),
\label{eq:Schroedinger}
\end{equation}
where $H_0 \equiv -\frac{1}{M}\frac{d^2}{dx^2}$ denotes the free
Hamiltonian, and $E_m\,(m=0,1,2,\cdots,m_c)$ are the eigenenergies
obtained from Eqs.~\eqref{eq:CCeq} in the box.
The energy-independent non-local HAL QCD potential $V(x,x')$ is
determined by demanding that Eq.~\eqref{eq:Schroedinger} reproduces the
NBS wave functions $\Psi(x;E_m)$.

With the (na\"ive) derivative expansion, the HAL QCD potential
$V(x,x')$ is expressed as
\begin{equation}
  V(x,x') 
  =
  \sum_{n=0}^\infty
  u_n(x)
  \left(
  \frac{\partial}{\partial x}
  \right)^n
  \delta(x-x').
\label{eq:NaiveDE}
\end{equation}
In this expansion, the lowest order term represents local
contribution, whereas the higher order terms yield non-locality
because of the derivatives.

\subsection{\label{sec:GDE}Generalized Derivative Expansion}

The na\"ive derivative expansion~\eqref{eq:NaiveDE} causes a problem
when it is applied to the Birse model.
Since the model involves the square-well and the $\delta$ functional
coupling potentials, NBS wave function~\eqref{eq:NBS-Birse} is not
smooth.
Then we rewrite the Schr\"odinger Eq.~\eqref{eq:Schroedinger} as
\begin{equation}
  \left(-H_0 + E_m\right)
  \Psi(x; E_m)
  =
  \sum_{n=0}^{\infty}
  u_n(x)
  \frac{d^n \Psi}{dx^n}(x; E_m)
  \label{eq:RHSNaive}
\end{equation}
to find that, on the right hand side, $\frac{d^n\Psi}{dx^n}(x;E_m)$
are singular at $x=0, \pm R$, involving derivatives of $\delta(x)$.

We generalize the derivative expansion to avoid this problem.
We replace $\delta$ functional kernel in the na\"ive
expansion~\eqref{eq:NaiveDE} by a Gaussian kernel
\begin{equation}
  \delta_\rho(x-x')
  \equiv
  \frac{\exp\left(-(x-x')^2/\rho^2\right)}{\sqrt{\pi}\rho},
\label{eq:Gaussian-kernel}
\end{equation}
where an arbitrary scale parameter $\rho$ is introduced.
We will refer to the scale as the {\em Gaussian expansion scale}.
This replacement leads to the generalized derivative expansion:
\begin{align}
  V(x,x')
  &=
  \sum_{n=0}^\infty
  v_n^{(\rho)}(x)
  \left(\frac{\partial}{\partial x}\right)^n
  \delta_\rho(x - x')   \label{eq:GeneralizedDEx} \\
  &=\sum_{n=0}^\infty
  v_n^{(\rho)}(x)
  \frac{1}{\rho^n}
  H_n\left(-\frac{x-x'}{\rho}\right)
  \delta_\rho(x - x'),   \nonumber
\end{align}
where, in the second line, the derivatives are replaced by the
Hermite polynominals $H_n(x) \equiv (-1)^n e^{x^2} (d/dx)^n e^{-x^2}$.
For notational simplicity, we define the smoothed wave function
$\Phi_\rho(x;E_m)$ according to
\begin{equation}
  \Phi_\rho(x;E_m)
  \equiv
  \int dx'\,
  \delta_\rho(x - x')
  \Psi (x';E_m),
  \label{eq:smearedWF}
\end{equation}
to rewrite the Schr\"odinger Eq.~\eqref{eq:Schroedinger} as
\begin{equation}
  \left(-H_0 + E_m\right)
  \Psi (x; E_m)
  =
  \sum_{n=0}^{\infty}
  v_n^{(\rho)}(x)
  \frac{d^n \Phi_\rho}{dx^n}(x; E_m).
  \label{eq:smearing-schroedinger}
\end{equation}
Unlike in Eq.~\eqref{eq:RHSNaive}, the right hand side does not
involve any singularity, since $\Phi_\rho(x;E_m)$ is smooth
everywhere.

The new expansion~\eqref{eq:GeneralizedDEx} is a natural
generalization of the na\"ive expansion~\eqref{eq:NaiveDE}, since
$\delta_\rho(x)$ is reduced to $\delta(x)$ in the $\rho \to + 0$
limit.
In  Appendix~\ref{sec:A-GDE-ms}, we  give a  proof that  $V(x,x')$ can
always  be  expanded   as  Eq.~\eqref{eq:GeneralizedDEx}.

Readers should not confuse the replacement
$\Psi(x)\to\Phi_\rho(x)$ with the smearing of interpolating fields,
which is often used in lattice QCD calculations.  
The aim in introducing the Gaussian kernel $\delta_{\rho}(x - x')$ is
a different parameterization of the non-local potential $V(x,x')$ such
that it can be applied to unsmooth wave functions.
Keeping that in mind, it is clear that the smoothed wave function
$\Phi_\rho(x)$ should only appear on the right-hand side of
Eq.~\eqref{eq:smearing-schroedinger}, while it is the original wave
function $\Psi(x)$ that appears on the left-hand side.

We make two more technical modifications here.
Since we restrict ourselves to the parity-even sector, the Gaussian
kernel $\delta_\rho(x-x')$ is projected onto the even-parity
subspace by projection operator $\mathbb{P}^{(+)}$ according to
\begin{align}
\begin{split}
  W_{\rho}(x,x')
  &\equiv \mathbb{P}^{(+)} \delta_{\rho}(x-x') \mathbb{P}^{(+)} \\
  &= \frac1{2}\left[ \delta_{\rho}(x-x') + \delta_{\rho}(x+x') \right].
\end{split}
  \label{eq:W_rho}
\end{align}
In addition, we replace the $x$-derivative $\partial/\partial x$ by
the $x^2$-derivative,
\begin{equation}
  D_x
  \equiv
  \frac{\partial}{\partial     (x^2)}.
  \label{eq:D_x}
\end{equation}
The replacement~\eqref{eq:D_x} is necessary to avoid power divergence
in $v_n^{(\rho)}(x)$ at $x=0$, which is caused by the fact that
the odd-order x-derivatives of the smoothed wave function
$\Phi_\rho(x;E)$ vanishes at $x=0$ at any energy.
Since the conventional derivatives $\partial_x, (\partial_x)^2,
\cdots, (\partial_x)^N$ and $D_x, D_x^2, \cdots, D_x^N$ are related by
an invertible linear transformation for $0 < |x| < \infty$, the
replacement~\eqref{eq:D_x} simply corresponds to a rearrangement of
the expansion coefficients.
(See Appendix~\ref{sec:need-for-the-replacement} for detail.)
Thus, the modification~\eqref{eq:D_x} affect neither the physical
observables nor the HAL QCD potentials.

Let us comment on some important features of the generalized
derivative expansion, for application to other systems, including
lattice QCD.
Besides our original purpose of smoothing unsmooth wave functions, it
has advantage to the na\"ive expansion in two more points.
(1) As is seen in the second expression in
Eq.~\eqref{eq:GeneralizedDEx}, derivatives can be replaced by the
Hermite polynominals. In this way one can alternatively use numerical
integration, which is more stable than numerical derivative in many
cases.
(2) The choice of the expansion scale $\rho$ shall determine the rate
of convergence in the generalized expansion, and can be chosen
arbitrarily.  Thus, we expect that it is possible to improve the
convergence without additional computational cost by properly choosing
the scale.

\subsection{\label{sec:DETERMINATION}Potential Determination}

The final expression of our expansion is
\begin{equation}
  V^{(N)}(x,x')
  = \sum_{n=0}^N v_n^{(\rho,N)}(x) D_x^n \, W_\rho (x,x'),
\label{eq:GeneralizedDE-truncation}
\end{equation}
where the summation over $n$ is truncated at finite order $N$,
assuming that the higher order contributions are negligible in
describing scattering phenomena for $E<\Delta$.

Now we take the $N+1$ lowest-lying energy levels in a finite box, and
arrange the Schr\"odinger Eqs. for these energies in a matrix form as
\begin{widetext}
\begin{equation}
\begin{pmatrix} 
(E_0-H_0)\Psi(x;E_0) \\ (E_1-H_0)\Psi(x;E_1) \\ \vdots \\ (E_N-H_0)\Psi(x;E_N)
\end{pmatrix}
=
\begin{pmatrix}
\Phi_\rho(x;E_0) & D\Phi_\rho(x;E_0) & \cdots & D^N\Phi_\rho(x;E_0) \\
\Phi_\rho(x;E_1) & D\Phi_\rho(x;E_1) & \cdots & D^N\Phi_\rho(x;E_1) \\
   \vdots   &    \vdots    & \ddots &    \vdots      \\
\Phi_\rho(x;E_N) & D\Phi_\rho(x;E_N) & \cdots & D^N\Phi_\rho(x;E_N)
\end{pmatrix}
\begin{pmatrix}
v_0(x) \\ v_1(x) \\ \vdots \\ v_N(x)
\end{pmatrix}.
\label{eq:regular-part}
\end{equation}
\end{widetext}
To determine the HAL QCD potential,
matrix inversion is performed point-by-point to solve
Eq.~\eqref{eq:regular-part} for $(v_0(x),v_1(x),\cdots,v_N(x))^T$.
The energy levels are understood to be arranged in the ascending order,
such that $E_0 < E_1 < \cdots < E_N$.

Due to the contact interaction containing $\delta(x)$ in
Eq.~\eqref{eq:CCeq}, $H_0\Psi(x;E_m)$ on the left hand side of
Eq.~\eqref{eq:regular-part} has a $\delta$ functional singularity at
$x=0$.
Accordingly, each of the coefficients $v_n(x)$ also has a ``singular''
component, which needs separate treatment in numerical calculation.
We decompose the coefficients into the singular part and the
``regular'' part $\tilde{v}_n$ as
\begin{equation}
  v_n(x)=\tilde{v}_n(x)+g_n \delta(x).
  \label{eq:vn-decomposition}
\end{equation}
We then integrates Eq.~\eqref{eq:regular-part} in the interval
$-\epsilon < x < +\epsilon$, and takes the $\epsilon \to +0$ limit
thereafter.  The weight constants $g_n$ are determined by solving the
following matrix equation:
\begin{widetext}
\begin{equation}
  \begin{pmatrix}
    (2/M)\Psi'(x=+0;E_0)    \\    (2/M)\Psi'(x=+0;E_1)    \\    \vdots
    \\ (2/M)\Psi'(x=+0;E_N)
  \end{pmatrix}
  =
  \begin{pmatrix}
    \Phi_\rho(x;E_0) & D\Phi_\rho(x;E_0) & \cdots & D^N\Phi_\rho(x;E_0) \\
    \Phi_\rho(x;E_1) & D\Phi_\rho(x;E_1) & \cdots & D^N\Phi_\rho(x;E_1) \\
    \vdots   &    \vdots    & \ddots &    \vdots      \\
    \Phi_\rho(x;E_N) & D\Phi_\rho(x;E_N) & \cdots & D^N\Phi_\rho(x;E_N)
  \end{pmatrix}_{x=0}
  \begin{pmatrix}
    g_0 \\ g_1 \\ \vdots \\ g_N
  \end{pmatrix}.
\end{equation}
\end{widetext}

\subsection{Scattering Phase Shift}

Let us consider the Lippmann-Schwinger equation
\begin{equation}
  \ket{\Psi}=\ket{k}+\frac 1{E-H_0+i\epsilon}V\ket{\Psi},
  \label{eq:LSeq1}
\end{equation}
where $\ket{k}$ denotes a plane wave with asymptotic momentum $k\equiv
\sqrt{ME}$.   By   inserting  the   completeness  relation   $\int  dx
\ket{x}\bra{x}=\bm{1}$, the Lippmann-Schwinger equation in
the (1+1 dimensional) coordinate space is written as
{
\abovedisplayskip=5pt
\belowdisplayskip=0pt
\begin{equation}
  \Psi(x) = \phi(x)+\iint dx'dx''\, G(x,x')V(x',x'')\Psi(x''),
  \label{eq:LSeq2}
\end{equation}
}
{
\abovedisplayskip=0pt
\belowdisplayskip=10pt
\begin{align}
  \Psi(x) &\equiv \Braket{x|\Psi}, \nonumber \\
  \phi(x) &\equiv \Braket{x|\phi} = (2\pi)^{-1/2} \cos kx, \\
  G(x,x') &\equiv \Braket{x|\frac 1{E-H_0+i\epsilon}|x'} 
          = \frac{-iM}{2k} e^{ik|x-x'|}, \nonumber \\
  V(x',x'')&\equiv \Braket{x'|V|x''}. \nonumber
\end{align}
}
We use the non-local potential in
Eq.~\eqref{eq:GeneralizedDE-truncation} in place of $V(x',x'')$.
Scattering phase shift is extracted from the solution of
Eq.~\eqref{eq:LSeq2} based on the asymptotic behavior in
Eq.~\eqref{eq:as-behavior}.

\section{\label{sec:RESULTS}Numerical Results}

\subsection{\label{sec:RESULTS-TBC}Parameter Set and Boundary Condition}

We employ the same parameter set as given in Ref.~\cite{Birse:2012ph}, i.e.,
$MV_0=1/R^2,\,M\Delta=6/R^2,\,Mg=6/R$, and we take $R=1$ and $M=1$.
A single bound state is found at $E=E_0\simeq-33.7$ with this
parameter set.
We take large enough spatial volume such that the bound state energy
is almost insensitive to the finite volume correction.

We solve the coupled channel Eqs.~\eqref{eq:CCeq} analytically in a
finite box of $-L<x<+L$ under the twisted boundary conditions (TBC)
\begin{align}
\label{eq:TBC}
\begin{split}
  \psi_{0,1} (x+2L) &= e^{i\theta} \psi_{0,1} (x), \\
  \psi_{0,1}^* (x)  &= \psi_{0,1} (-x), 
\end{split}
\end{align}
with $L=10$ and $\theta=\pi/2$.
The second condition implies that the real part of the solution is
parity-even, whereas the imaginary part is parity-odd.
Hence, we use the  real part of the solution to  construct the HAL QCD
potentials in the parity-even sector.
In the followings, the symbols $\psi_0(x)$ and $\psi_1(x)$ are
exclusively used to indicate the real parts of the corresponding
solutions.
The exact form of the solution is given in Appendix~\ref{sec:A-BMS-1}.
We note that a technical problem arises around the boundary when the
conventional (anti-)periodic BC is imposed instead of the TBC.
(See Appendix~\ref{sec:A-GDE-TBC} for detail.)
%

We assign several values to the field admixture parameter $q$ and the
Gaussian expansion scale $\rho$, such that $|q|\leq 1.0$ and $0.1\leq
\rho \leq 0.7$, respectively.
In particular, we employ $q=+0.2$ and $\rho=0.5$ as reference values.
The other choices of these parameters lead to qualitatively same
results, except for the occurrence of the following technical
problems.
First, numerical calculation becomes unstable for $\rho \alt 0.1$ or
$N \agt 6$.  This is because the condition number associated with the
matrix inversion in Eq.~\eqref{eq:regular-part} gets too large to
determine the expansion coefficients precisely.
Moreover, there can arise another problem that the determinant of the
matrix becomes zero at some spatial points.  We find that the latter
case mainly matters with large $|q|$ values (typically with $q
\agt 3$), in the Birse model of the above parameters.
In Appendix.~\ref{sec:zero-determinant}, we will discuss the problem
in detail and point out a way to circumvent it by utilizing properties
of the generalized derivative expansion.

We make a comment on how the choice of the field admixture parameter
is reflected to the non-locality in this model.  Coupled-channel
Eqs.~\eqref{eq:CCeq} for $x\neq 0$ are expressed as
\begin{equation}
  \label{eq:non-locality-qdep}
    \left[
    -\frac1{M}\frac{d^2}{dx^2} + V_0(x) - E
    \right]
  \Psi_q(x)
  =
  q
  \left(
    V_0(x)-\Delta
  \right)
  \psi_1(x),
\end{equation}
since $\delta(x\neq 0)=0$.
It follows that, if $q=0$ is employed as the field admixture parameter,
the left-hand side of Eq.~\eqref{eq:non-locality-qdep} vanishes, and
the trivial local quantity $V_0(x)\delta(x-x')$ works as the HAL QCD
potential for $x\neq 0$.  The non-locality is accumulated at $x=0$ as
a sum of derivatives of the $\delta$ function.
With finite $q$, the potential becomes non-local even for $x\neq 0$.
Then we expect that smaller $|q|$ tends to result in a HAL QCD
potential with smaller non-locality, since the right-hand side of
Eq.~\eqref{eq:non-locality-qdep} is linear in $q$.

\subsection{\label{sec:WAVE-FUNCTIONS}NBS Wave Functions}

\begin{figure}[h]
  \centering
  \includegraphics[width=0.7\hsize, bb=50 50 410 292]{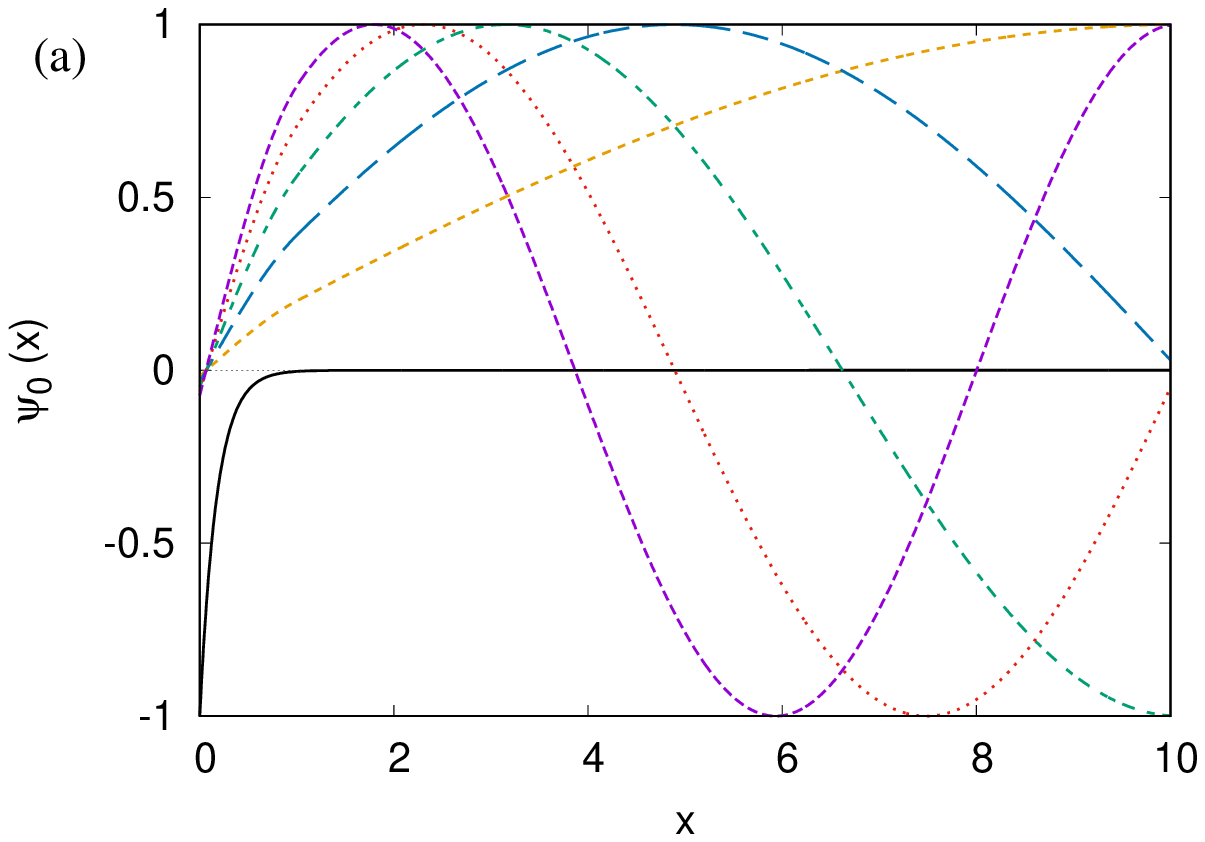}
  \includegraphics[width=0.7\hsize, bb=50 65 410 292]{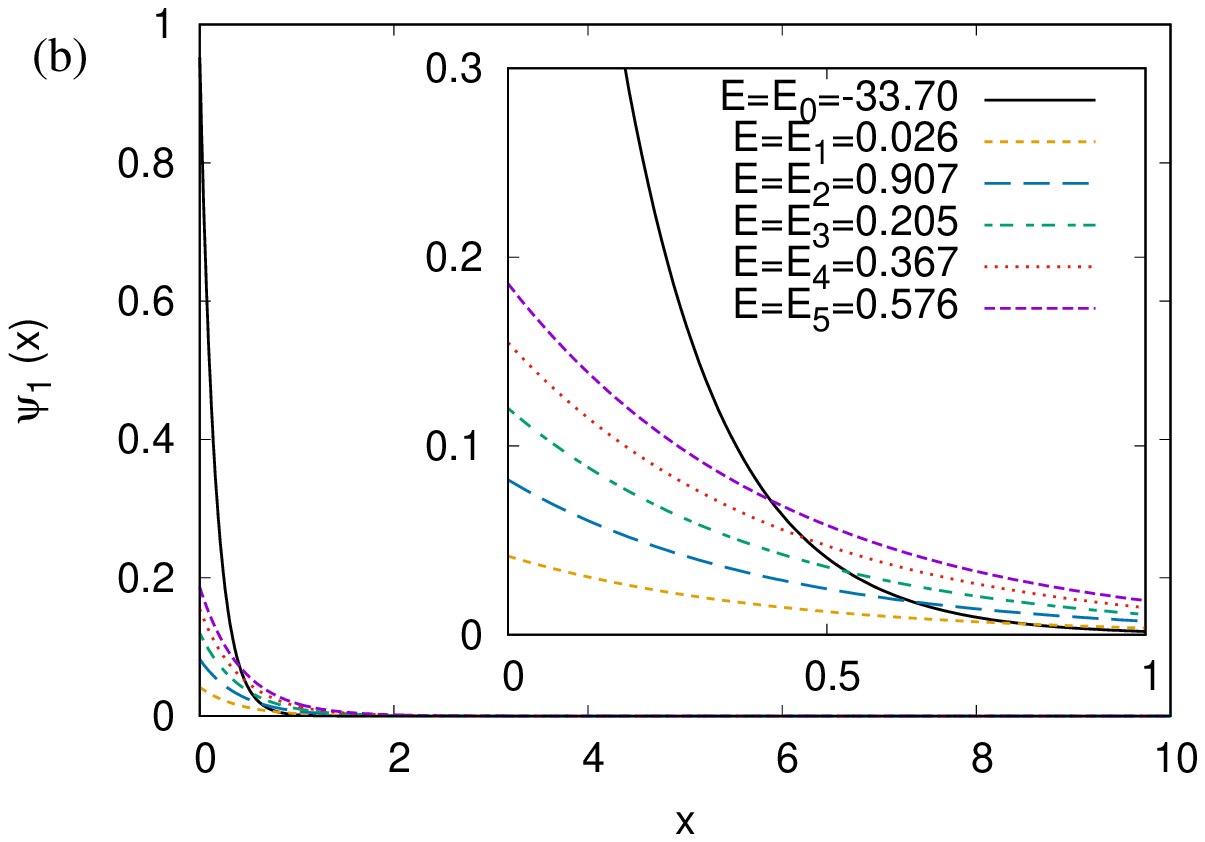}
  \caption{
    \label{fig:psi0psi1}
    Wave functions (a) $\psi_0(x)$ and (b) $\psi_1(x)$ for the six
    lowest-lying-energy states with the TBC.  }
\end{figure}

Figure \ref{fig:psi0psi1} shows the wave functions $\psi_0(x)$ and
$\psi_1(x)$ for the six lowest energy levels $E_0, E_1, \cdots, E_5$
with the TBC.
Since the overall normalization factor is irrelevant to the
determination of the potential, we take arbitrary normalization for
visibility.
Notice that the ground state with $E=E_0$ is a bound state and
$\psi_0(x;E_0)$ is localized around the origin. As already indicated
in Eq.~\eqref{eq:closed_suppressed}, the $\psi_1$ channel is closed
for $E<\Delta$.

\begin{figure}[h]
  \centering
  \includegraphics[width=0.8\hsize, bb=50 65 410 282]{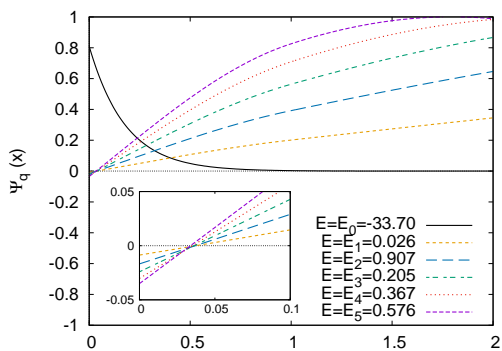}
  \caption{
    \label{fig:Psiq}
    Nambu-Bethe-Salpeter wave functions
    $\Psi_q(x)=\psi_0(x)+q\psi_1(x)$ with $q=+0.2$.  For better
    visibility, we inverted the sign of the NBS wave function
    for $E=E_0$ (since the overall normalization is irrelevant)
    and altered the $x$-range from Fig.~\ref{fig:psi0psi1}.
  }
\end{figure}

In Fig.~\ref{fig:Psiq}, we show the NBS wave functions $\Psi_q(x)$
with $q=0.2$ for the same energy levels.  For $E \geq E_1$, the NBS
wave functions have nodes at $x \simeq 0.035$.  (The nodal positions
are slightly energy-dependent.)
The nodal position is $q$ dependent, and no short-distance node is
observed with $q \gtrsim 0.5$.

\subsection{\label{sec:RESULTS-POT}Non-Local Potentials}

\begin{figure*}[p]
  \begin{center}
    \includegraphics[width=0.45\hsize]{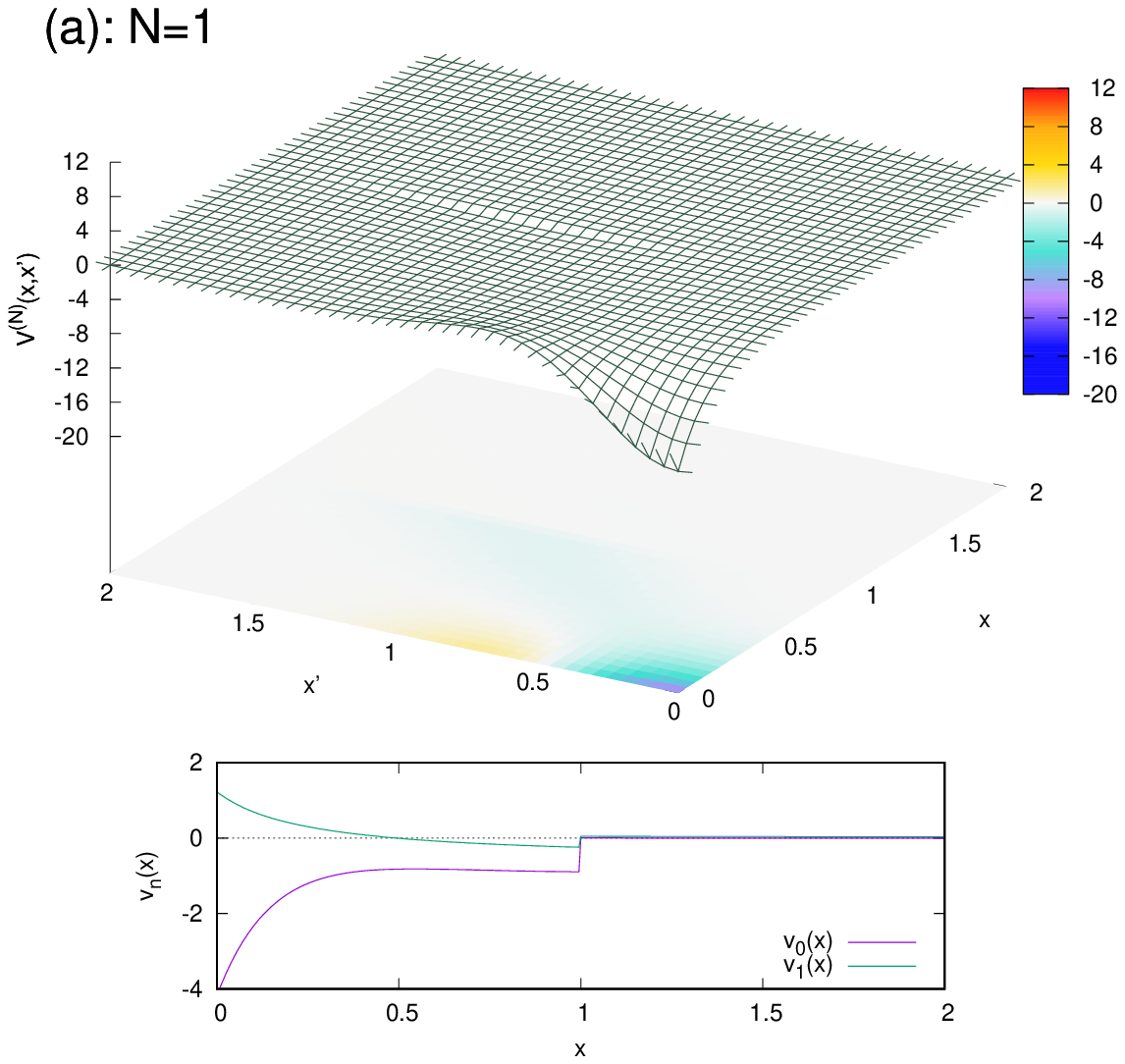}
    \includegraphics[width=0.45\hsize]{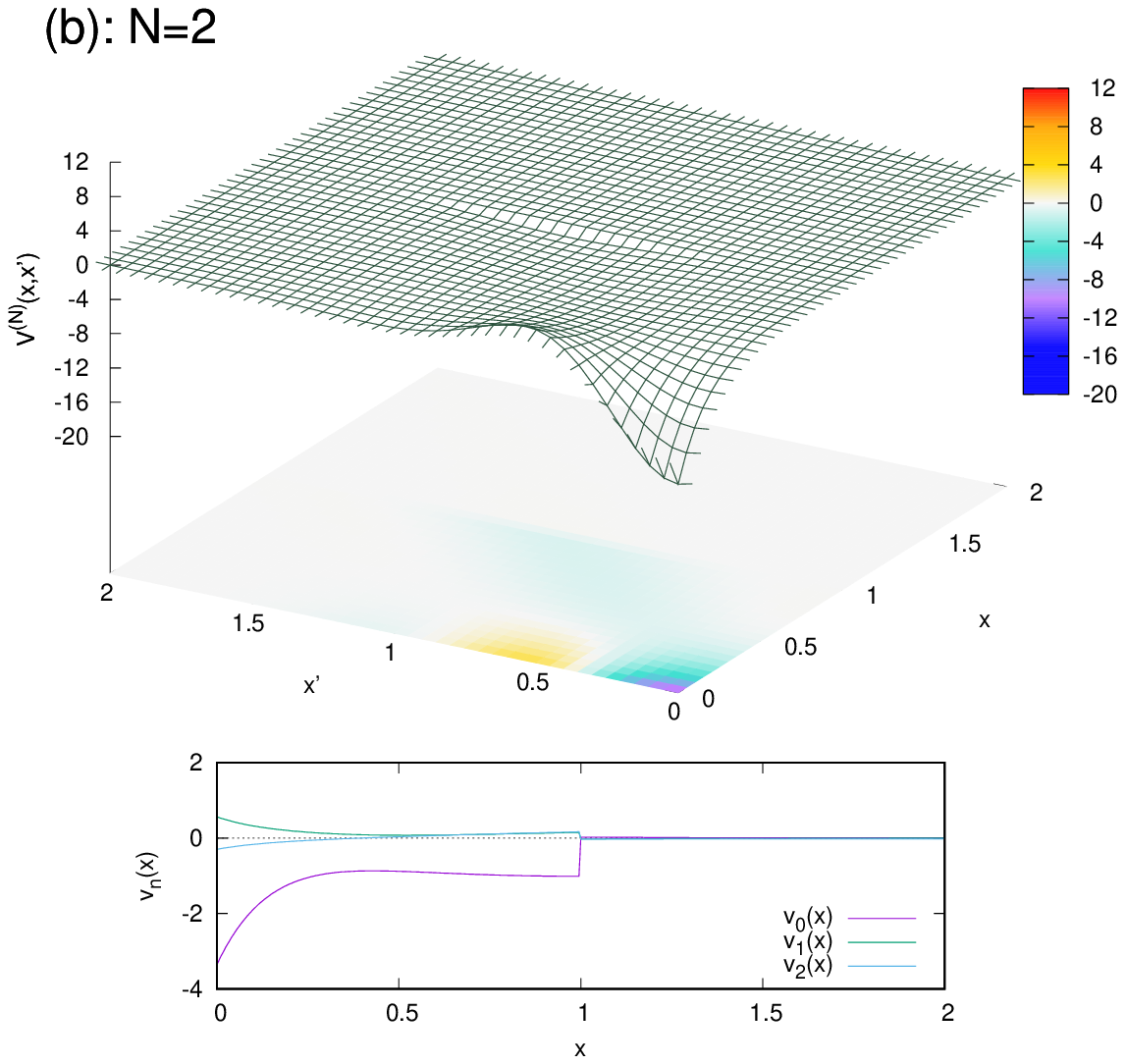}
    \includegraphics[width=0.45\hsize]{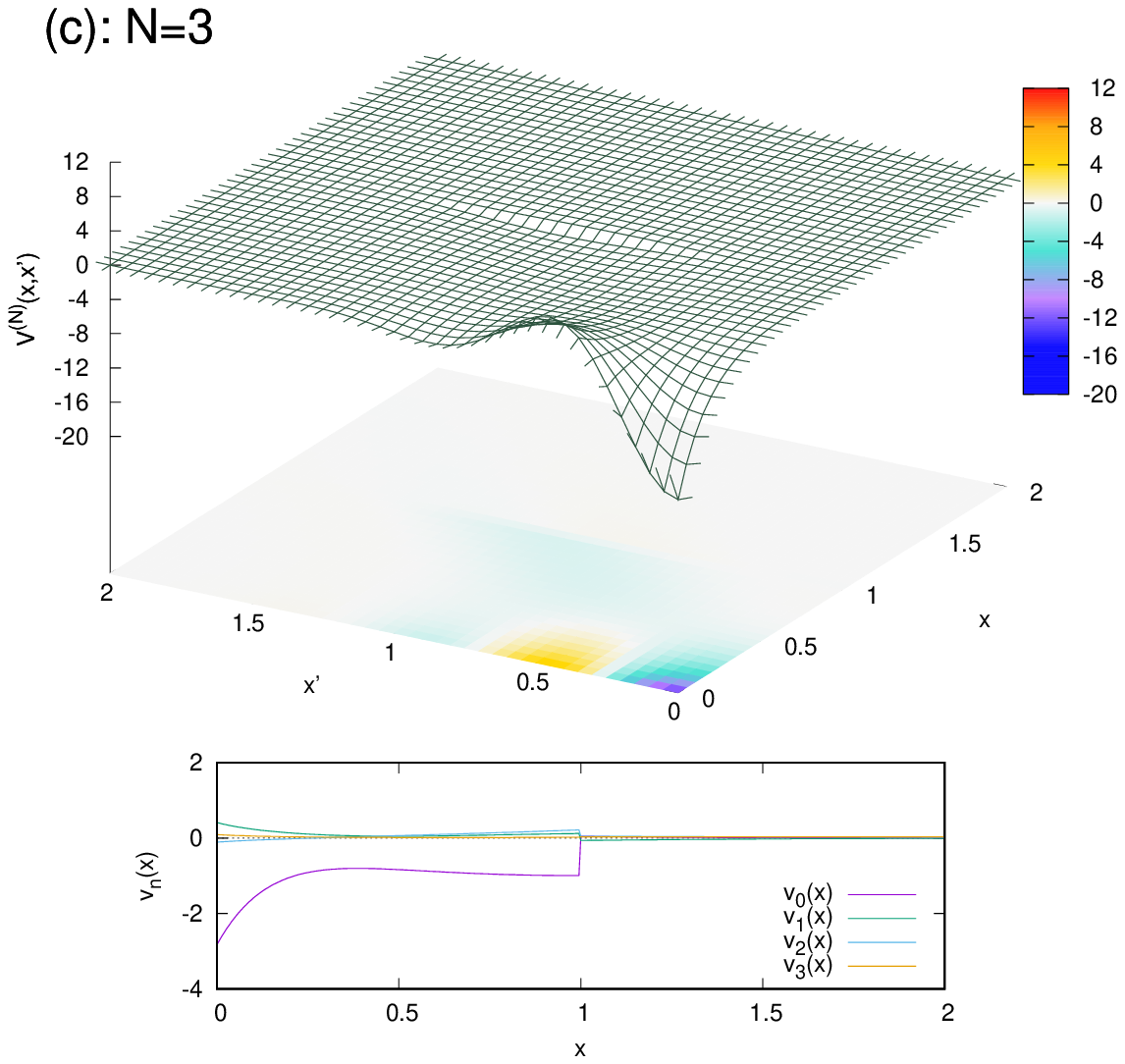}
    \includegraphics[width=0.45\hsize]{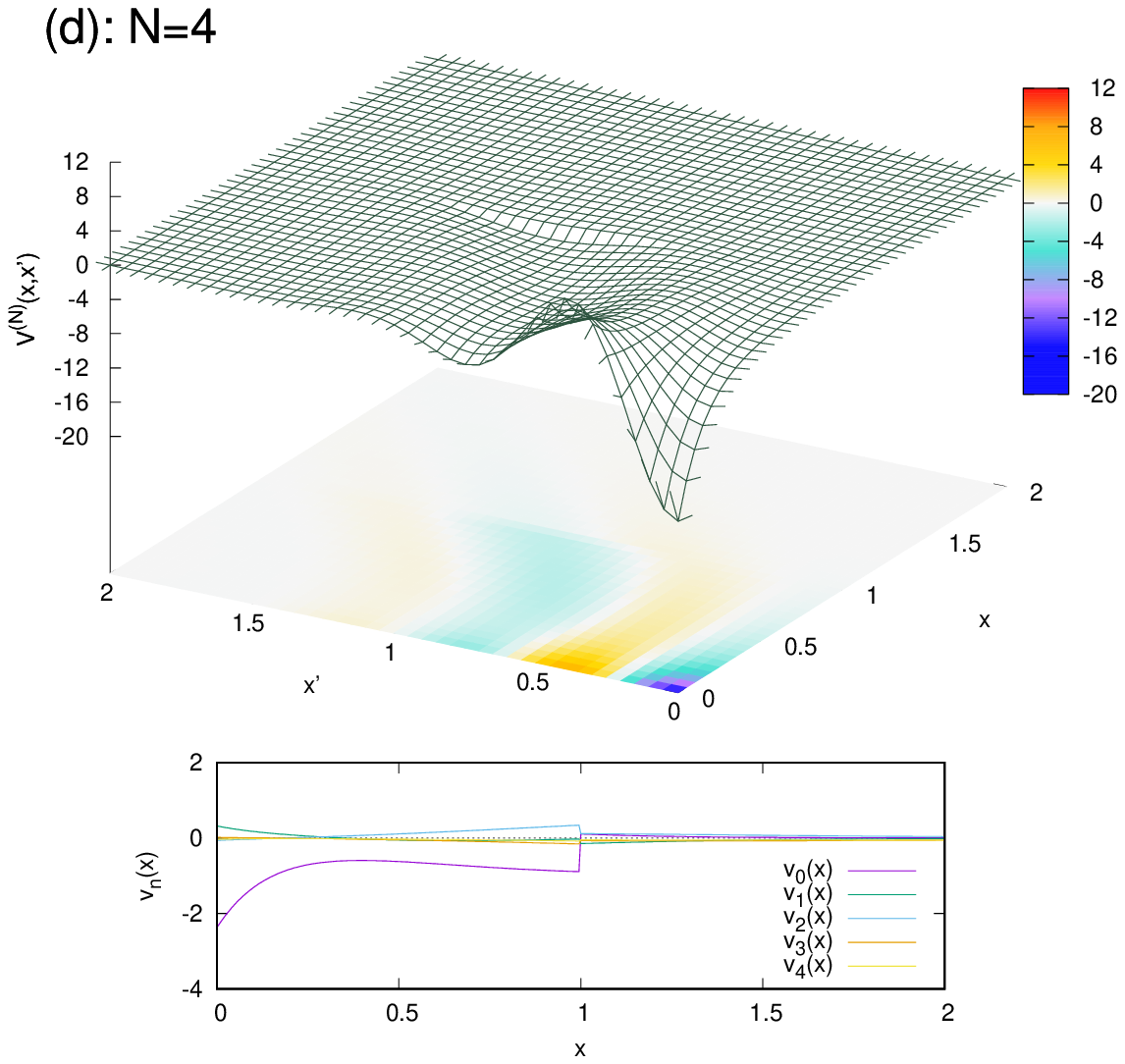}
    \includegraphics[width=0.45\hsize]{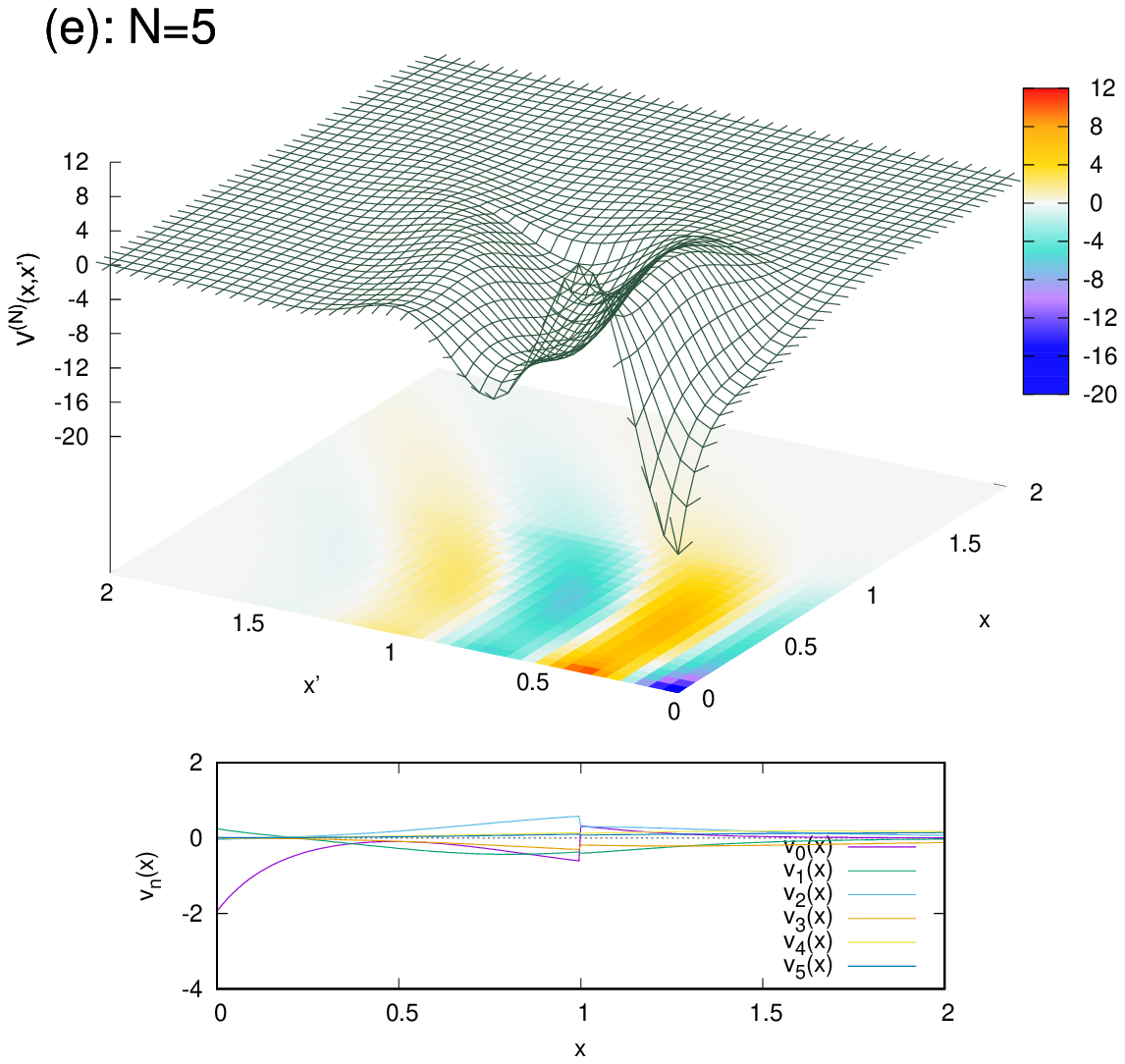}
  \end{center}
  \caption{
    \label{fig:HALQCDpot_and_coefficients}
    Regular part of the non-local HAL QCD potentials $V^{(N)}(x,x')$
    and their expansion coefficients $v_n(x)$ ($n=0,\cdots,N$) with
    $\rho=0.5$ and $q=0.2$.  Panels (a), (b), (c), (d), and (e) show
    the results of $N=1, 2, 3, 4,$ and $5$, respectively.
  }
\end{figure*}

In Fig.~\ref{fig:HALQCDpot_and_coefficients}, we show the regular part
of the HAL QCD potentials $V^{(N)}(x,x')$ and their expansion
coefficients $\tilde{v}_n(x)$ at truncation orders $N=1,2,3,4,$ and
$5$. 
The field admixture parameter and the Gaussian expansion scale are
fixed to $q=0.2$ and $\rho=0.5$, respectively.
The discontinuous structure seen in the potentials at $|x|=R=1$ is due
to the edge of the square-well potential in the Birse model.
The strength $g_n$ of their singular parts are summarized in
Table~\ref{table:gn}.

We observe that the functional structure of the non-local potentials
is not stable against the variation of $N$ as the magnitude of the
potential tends to become larger for larger $N$.
In principle, a potential is an off-shell object and variance in its
functional structure itself does not necessarily imply that the
expansion fails.

Unlike in Ref.~\cite{Birse:2012ph}, there is no repulsive behavior in the
small $|x|$ region.
In Ref.~\cite{Birse:2012ph}, the leading order potential of the na\"ive
derivative expansion~\eqref{eq:NaiveDE} is computed from a single NBS
wave function of the first excited state.
As we have seen in Fig.~\ref{fig:Psiq}, the NBS wave function of the
first excited state has a node at short distance.  At that point the
HAL QCD potential with the leading order derivative expansion
diverges, and it looks as if there were a repulsive core.
To avoid such superficial divergence, two or more NBS wave functions
should be picked up from the lowest energy without omission to solve
the matrix Eq.~\eqref{eq:regular-part} for $(v_0(x), \cdots,
v_N(x))^T$.
In this case, the condition $\Phi_{\rho}(x;E_i) = 0$ for some
combination of $x$ and $E_i$ does not necessarily mean that the
inversion of the whole matrix fails in solving
Eq.~\eqref{eq:regular-part}.  We can thus obtain a smooth potential
even at the nodal position.

\begin{figure}[h]
  \centering
  \includegraphics[width=0.8\hsize]{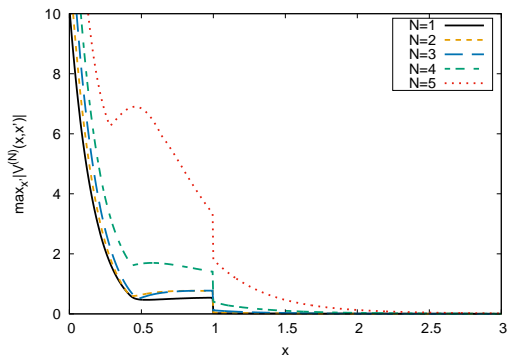}
  \caption{
    \label{fig:pot_profile}
    Function $f^{(N)}(x)$ with $\rho  = 0.5$ and $q = 0.2$. (See Eq.~\eqref{eq:maximum})
  }
\end{figure}

In Fig.~\ref{fig:pot_profile}, we plot the following quantity 
for $N=1,\cdots,5$ to see the range of the potentials:
\begin{equation}
  f^{(N)}(x)
  \equiv
  \underset{x'}{\mbox{max}}
  \left|V^{(N)}(x,x')\right|.
  \label{eq:maximum}
\end{equation}
It is clear that $f^{(N)}(x) \simeq 0$ for $x>3$.  In the calculation
of the phase shift from the Lippmann-Schwinger equation below, we
safely regard $V^{(N)}(x,x') = 0$ for $|x| > 4$.

\begin{table}[h]
  \caption{
    \label{table:gn}
    Singular part strength $g_n$ of the potentials in
    Fig.~\ref{fig:HALQCDpot_and_coefficients}.
  }
  \begin{tabular}{@{\hspace{0.5em}}c r @{\hspace{1.5em}} r @{\hspace{1.2em}} r @{\hspace{1em}} r @{\hspace{1em}} r @{\hspace{1em}} r @{\hspace{0.5em}}} \hline \hline
    $N$  &  $g_0$   &  $g_1$  &  $g_2$  &  $g_3$  &  $g_4$  &  $g_5$  \\ \hline
    $1$  & $-16.31$ & $ 5.58$ &         &         &         &         \\
    $2$  & $-12.26$ & $ 2.32$ & $-1.47$ &         &         &         \\
    $3$  & $ -9.46$ & $ 1.51$ & $-0.46$ & $ 0.48$ &         &         \\
    $4$  & $ -6.85$ & $ 1.00$ & $-0.19$ & $ 0.05$ & $-0.19$ &         \\
    $5$  & $ -4.20$ & $ 0.54$ & $-0.05$ & $-0.03$ & $ 0.04$ & $ 0.09$ \\ \hline \hline
  \end{tabular}
\end{table}

\subsection{\label{sec:RESULTS-PS}Scattering Phase Shift}
\subsubsection{Convergence of the derivative expansion ($N$ dependence)}
\begin{figure}[t]
  \centering
  \includegraphics[width=0.9\hsize]{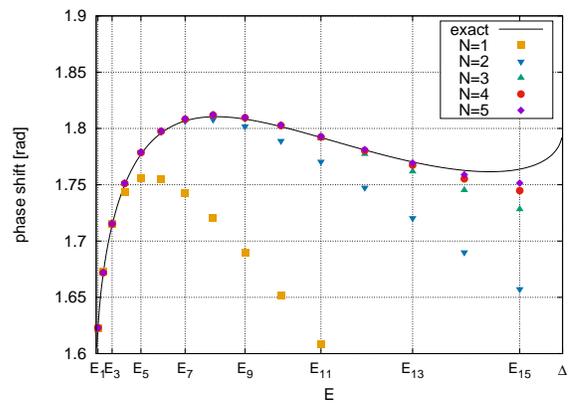}
  \caption{
    \label{fig:PhaseShift_Nchange}
    Scattering  phase  shift $\delta_N$  against  energy $E$  at
    $N=1,\,2,\,3,\,4,\,$ and  $5$, with  fixed $\rho=0.5$  and $q=0.2$.
    The  solid line  represents  the exact  values   extracted   from 
    the  analytic solution of the Birse model (similarly in 
    Fig.~\ref{fig:PhaseShift_qchange} and Fig.~\ref{fig:PhaseShift_rhochange}). 
    }
\end{figure}

In Fig.~\ref{fig:PhaseShift_Nchange}, we show the scattering phase
shifts extracted from the HAL QCD potentials with truncation orders
$N=1,\cdots,5$.  (Here we refer to these results as $\delta_N(E)$.)
In order to make discussion clear, we show $\delta_N(E)$ only at the
energies corresponding to the discrete eigenenergies of the Birse
model solutions in the box , i.e., $E=E_1,\cdots,E_{15}$.
The field admixture parameter and the expansion scale are fixed to
$q=0.2$ and $\rho=0.5$, respectively, so that the HAL QCD potentials
correspond to the ones shown in
Fig.~\ref{fig:HALQCDpot_and_coefficients}.
The results are compared to the exact values $\delta_{exact}(E)$
extracted from the analytic solution of the coupled-channel
Eqs.~\eqref{eq:CCeq}. (Refer to Appendix~\ref{sec:A-BMS-2} for the
explicit expression.)
We see that, at each truncation order $N$, $\delta_N$ agrees with
$\delta_{exact}$ at small energies, but deviates away at higher
energies.  However, the deviation tends to become smaller as $N$
increases.
This observation supports that the generalized derivative expansion
converges in terms of the phase shifts extracted from the non-local
potentials.

It is helpful to discuss the energy dependence in $\delta_N$ through
the following classification of the energy region.
First, we see that each of the results of $\delta_N$ in
Fig.~\ref{fig:PhaseShift_Nchange} shows excellent agreement with
$\delta_{exact}$ at discrete energies $E=E_1,\cdots,E_N$.
The agreement is ensured by construction, since the corresponding NBS
wave functions are used as input.
We also find that agreement holds in the intermediate intervals
$E_1<E<E_2,\cdots,E_{N-1}<E<E_N$.
We refer to the entire $E\le E_N$ interval as the ``{\em input
  region}''.
Secondly, we can see from Fig.~\ref{fig:PhaseShift_Nchange} that
agreement between $\delta_N$ and $\delta_{exact}$ extends beyond the
input region up to some point.
The agreement indicates that extrapolation to higher energy in terms 
of the phase shift is possible.
We refer to such a region with nontrivial agreement as the
``{\em extrapolation region}''.

In order to discuss the validity of truncation in the derivative
expansion, the length of the extrapolation region is of question.
From Fig.~\ref{fig:PhaseShift_Nchange}, we can see that at
$N=1,2,3,4,$ and $5$, the results show extrapolated agreement in the
intervals $E_1<E<E_3$ ($0.03 \Delta$), $E_2<E<E_8$ ($0.24 \Delta$),
$E_3<E<E_{11}$ ($0.45 \Delta$), $E_4<E<E_{12}$ ($0.51 \Delta$), and
$E_5<E<E_{13}$ ($0.58 \Delta$), respectively, where the numbers in the
parentheses are their lengths.
This observation reconfirms the convergence of the expansion, as
higher truncation order results in the extension of not only the
input region, but of the extrapolation region, i.e., more robust
extrapolation to higher energy.

Note that our phase shifts satisfy $\delta(E=0)=\pi/2$ in the 1+1
dimensional space-time with a single bound state, which is contrasted
to $\delta_l(E=0)-\delta_l(E=\infty)=n_l\pi$ in 1+3 dimension, with
$n_l$ being the number of bound states of angular momentum $l$.

\subsubsection{Interpolating field dependence ($q$ dependence)}
\label{sec:q-dependence}

\begin{figure*}
  \begin{minipage}{0.49\hsize}
    \centering
    \includegraphics[width=0.85\hsize]{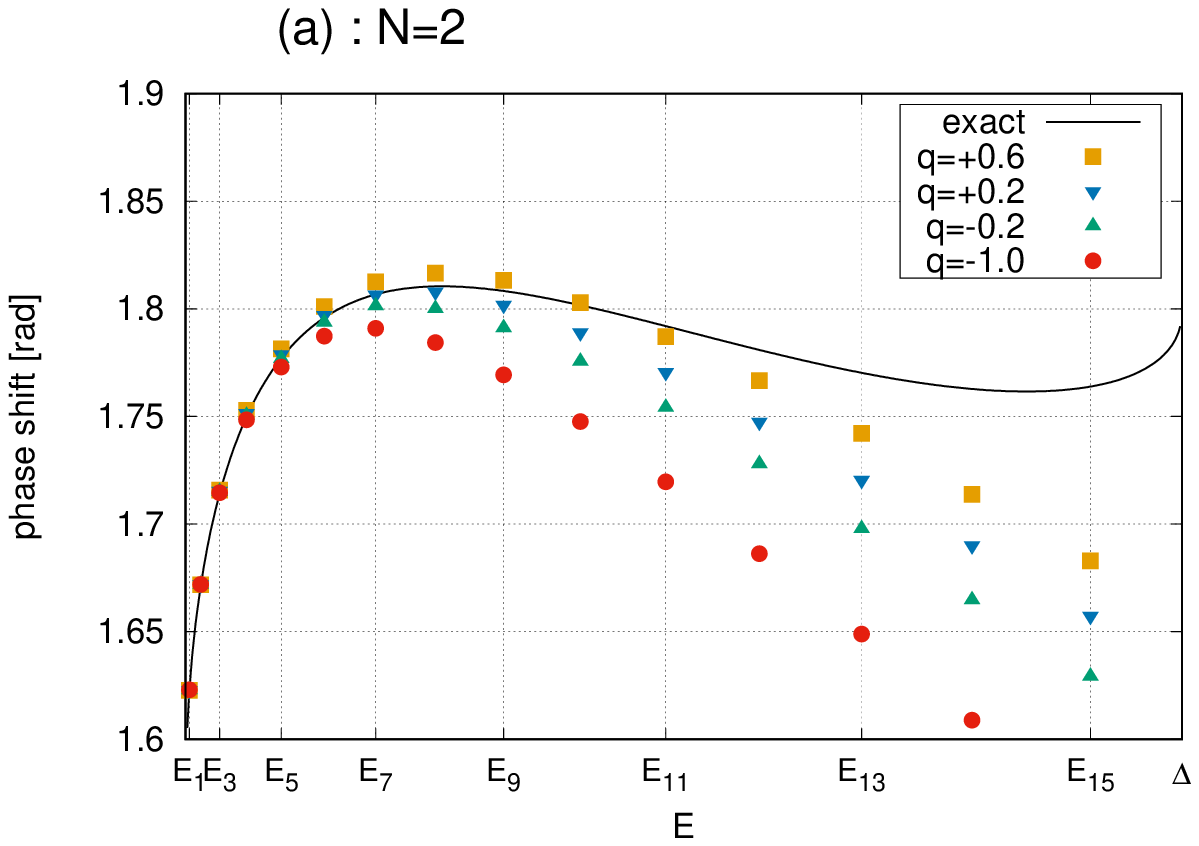}
  \end{minipage}
  \begin{minipage}{0.49\hsize}
    \centering
    \includegraphics[width=0.85\hsize]{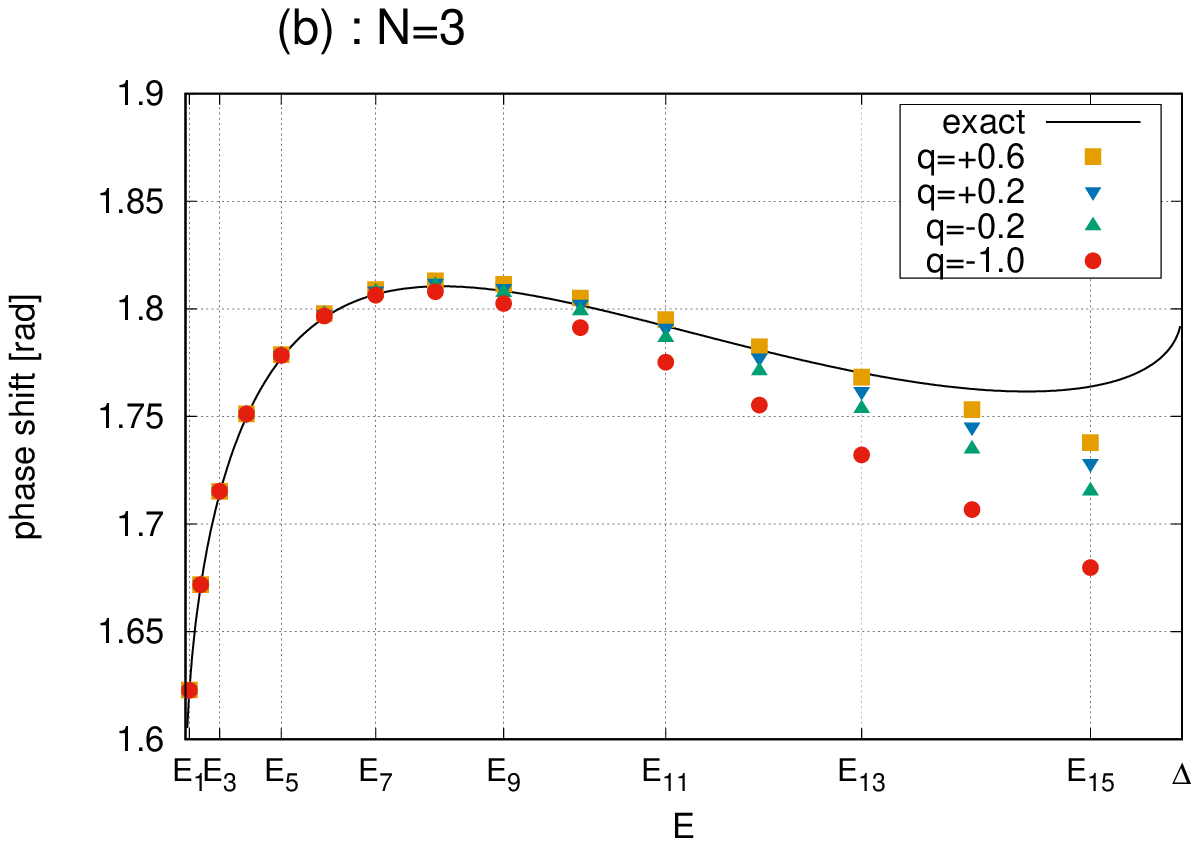}
  \end{minipage}
  \caption{
    \label{fig:PhaseShift_qchange}
    Scattering phase shift for $q=0.6,\,0.2,\,-0.2,\, -1.0$ with
    fixed $\rho = 0.5$ at (a) $N=2$ and (b) $N=3$.
  }
\end{figure*}

To discuss the dependence on the choice of interpolating fields, we vary
the field admixture parameter as $q = -1.0, -0.2, +0.2, +1.0$, while
the Gaussian expansion scale is fixed to $\rho = 0.5$.
Figure~\ref{fig:PhaseShift_qchange} shows the $q$ dependence of the
scattering phase shift at truncation orders $N=2$ and $N=3$.

At $N=2$, clear $q$ dependence can be seen; for $E>E_5$, the phase
shifts behave in a distinguishable manner.
The $q$ dependence is smaller at $N=3$, where the variance in
$\delta_N(E)$ against $q$ is smaller than at $N=2$ and all of the
results show better agreement with the exact curve.
By further increasing $N$ (despite the corresponding figures being
omitted to save the space), we observe increasingly smaller $q$
dependence in the phase shifts.
This implies that the generalized derivative expansion converges
regardless of the choice of interpolating fields.

To look into how the convergence is affected by the variance in $q$,
we first look at the results with $q=-1.0$.
The extrapolation regions with $q=-1.0$ correspond to $E_2<E<E_5$ at
$N=2$, and $E_3<E<E_8$ at $N=3$.  
%

\begin{figure*}
  \begin{minipage}{0.49\hsize}
    \centering
    \includegraphics[width=0.85\hsize]{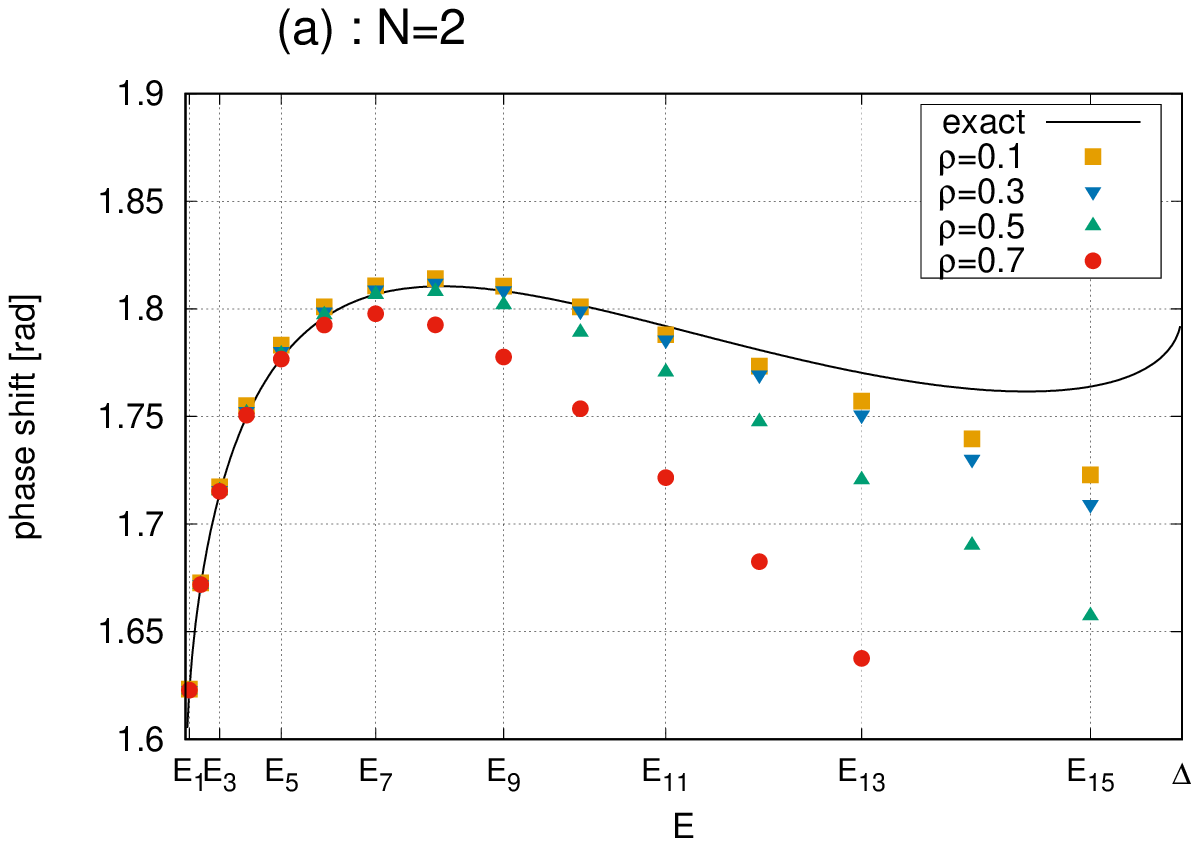}
  \end{minipage}
  \begin{minipage}{0.49\hsize}
    \centering
    \includegraphics[width=0.85\hsize]{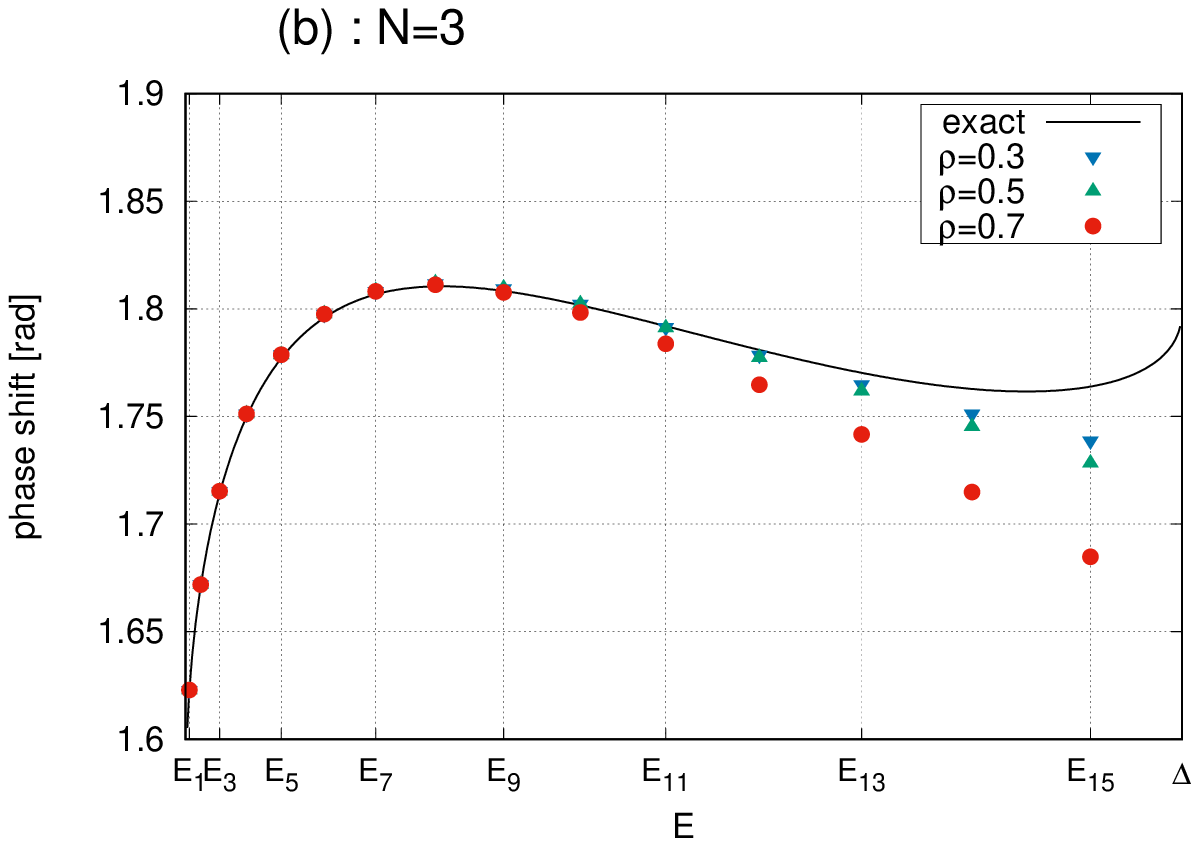}
  \end{minipage}
  \caption{
    \label{fig:PhaseShift_rhochange}
    Scattering phase shift for (a) $\rho=0.1,\,0.3,\,0.5,\,0.7$ at
    $N=2$ and (b) $\rho=0.3,\,0.5,\,0.7$ at $N=3$. In both panels the
    field admixture parameter is fixed to $q=0.2$.
  }
\end{figure*}

\noindent
The extension of the extrapolation
region reconfirms the above-mentioned convergence of the expansion.
In comparison to these results, we take the case of $q=+0.2$ to find
that the extrapolation regions are $E_2<E<E_8$ at $N=2$, and
$E_3<E<E_{11}$ at $N=3$.
It is observed that the lengths of the extrapolation regions are
comparable between the results with $(q,N)=(+0.2,\,2)$ and
$(q,N)=(-1.0,\,3)$ despite the different truncation orders.
It indicates that the interpolating field corresponding to $q=+0.2$
results in better convergence than the one corresponding to $q=-1.0$.
We therefore observe that the convergence of the generalized
derivative expansion is improved by properly choosing interpolating
fields.
Among the four $q$ values employed here, $q=+0.2$ seems to be the best
choice.

\subsubsection{Dependence on the Gaussian expansion scale $\rho$}

It is expected that the convergence of the generalized derivative
expansion can be improved by setting the expansion scale to the
intrinsic non-locality size of HAL QCD potentials, as well as tuning
the choice of interpolating fields.
To see this, we vary the expansion scale $\rho$ while the field
admixture is fixed to $q=+0.2$. In
Fig.~\ref{fig:PhaseShift_rhochange}(a), we show the phase shifts
obtained from HAL QCD potentials with $\rho=0.1, 0.3, 0.5$, and $0.7$
at truncation order $N=2$.
Similarly, Fig.~\ref{fig:PhaseShift_rhochange}(b) shows the phase
shifts with $\rho=0.3, 0.5, 0.7$ at $N=3$.

Let us compare the result with $(\rho,N)=(0.3,\,2)$ in
Fig.~\ref{fig:PhaseShift_rhochange}(a) and that with
$(\rho,N)=(0.7,\,3)$ in Fig.~\ref{fig:PhaseShift_rhochange}(b).
In the former case, the extrapolation region extends to $E_2<E<E_{10}$,
whereas that in the latter case corresponds to $E_3<E<E_{10}$.
Since both cases with different truncation orders allow for
extrapolation to comparable extrapolation regions, we conclude that a
proper choice of the Gaussian expansion scale can improve the
convergence of the generalized derivative expansion.

\subsection{Criteria for Convergence}

\begin{figure}[h]
  \begin{center}
    \includegraphics[width=0.95\hsize, bb=50 55 410 302]{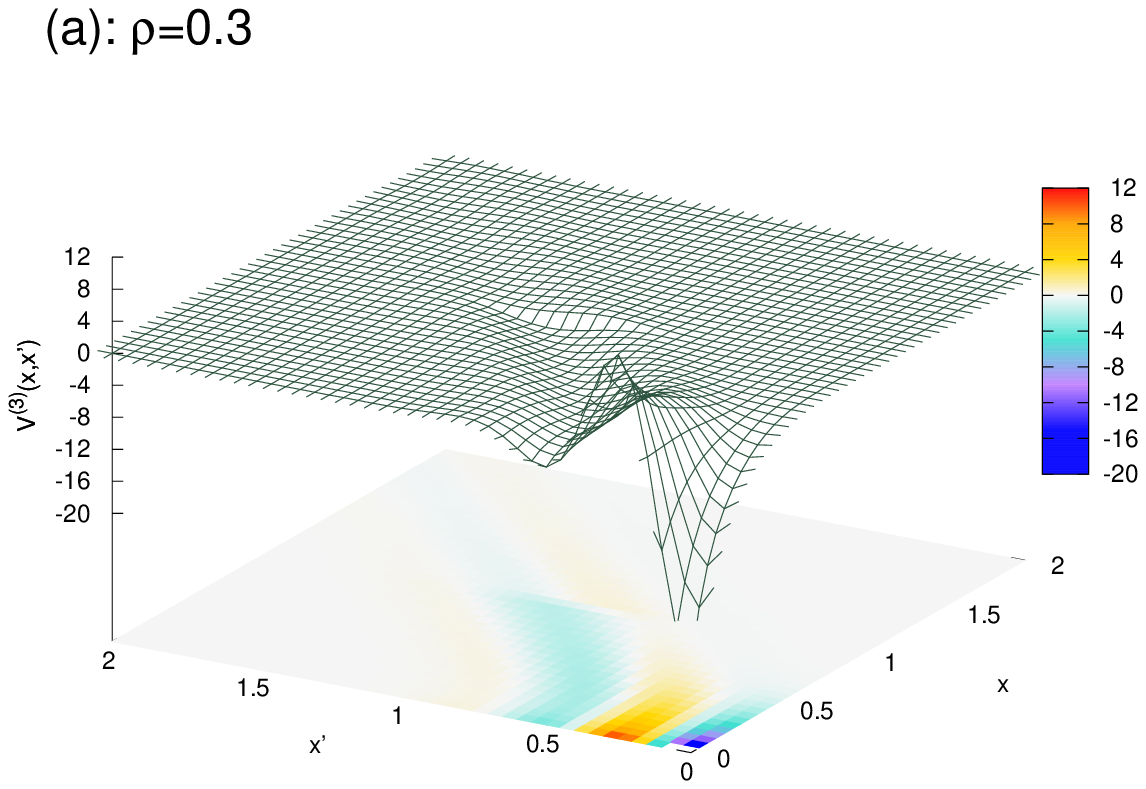}
    \includegraphics[width=0.95\hsize, bb=50 55 410 302]{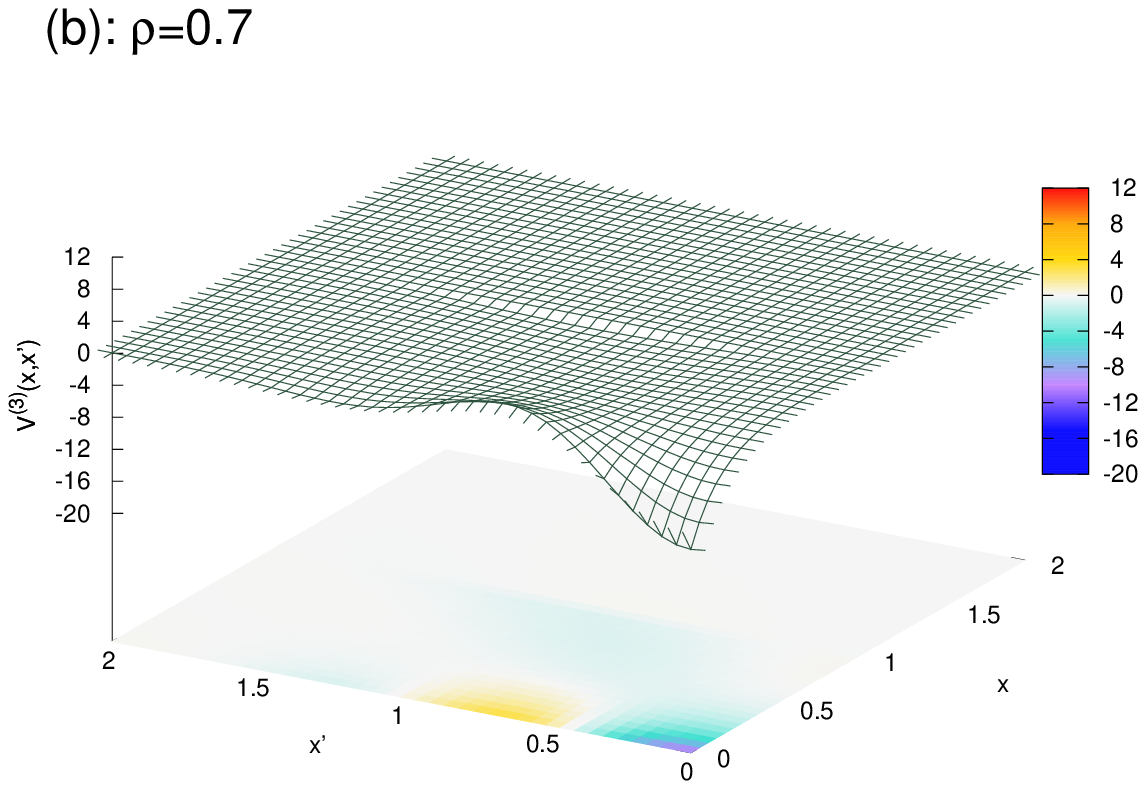}
  \end{center}
  \caption{
    \label{fig:potential_rho-dep}
    Non-local potentials $V^{(N)}(x,x')$ with (a) $\rho=0.3$ and (b)
    $\rho=0.7$, at truncation order $N=3$. The field admixture
    parameter is fixed to $q=+0.2$ in both panels.  
  }
\end{figure}

The  non-local  potentials with  $\rho=0.3$  and  $0.7$ are  shown  in
Fig.~\ref{fig:potential_rho-dep} with $N=3$ and $q=+0.2$ being fixed.
Although their structures seem to be quite different, these two
potentials reproduce the scattering phase shift quite well in the
energy region $E \le E_9$ (See
Fig.~\ref{fig:PhaseShift_rhochange}(b)).
Note that $E=E_9$ belongs to the extrapolation regions for both of
these two parameter sets.
%

\begin{figure}[h]
  \begin{center}
    \includegraphics[width=0.85\hsize]{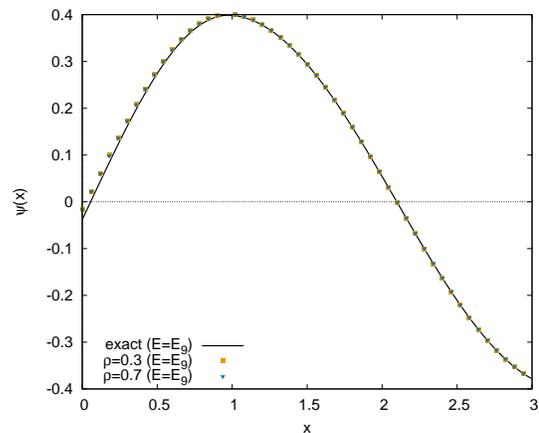}
  \end{center}
  \caption{
    \label{fig:WaveFunction_ShortRange}
    NBS wave  functions  at  $E=E_9$   obtained  with  the  two  non-local
    potentials  in  Fig.~\ref{fig:potential_rho-dep} (i.e.,  the  ones
    with  $\rho=0.3$ and  $\rho=0.7$, with  fixed $q=+0.2$  and $N=3$)
    together with the  analytic solution of the NBS  wave function for
    comparison.  }
\end{figure}

It  may  be  of  interest  how   the  wave  functions  behave  in  the
extrapolation region.
We show the NBS wave functions at $E=E_9$ obtained from the two
potentials together with the exact one in
Fig.~\ref{fig:WaveFunction_ShortRange}.  We see that, while they agree
at long distance, they show small deviation from the exact one at
short distance of $x \alt 0.5$.
The short-distance deviation is natural because the NBS wave function
at $E=E_9$ is not used as input to construct these potentials, although
the agreement in the phase shift ensures the correct long-distance
behavior of the NBS wave functions.

The difference in the structures between the two potentials in
Fig.~\ref{fig:potential_rho-dep} indicates that the HAL QCD potential
may not be uniquely determined as far as we use the NBS wave functions
for a restricted energy region.
This may be also the case even if we use all the NBS wave functions in
the energy region $E \le \Delta$.
From these considerations, we learn that the stability in non-local
potentials is too strict as a criterion for the convergence of the
(generalized) derivative expansion.  Physical observables, such as the
phase shift will be more useful for that purpose.

\section{\label{sec:CONCLUSION}CONCLUSION}

We have investigated general properties of the non-local potentials
which HAL QCD collaboration introduced as wave function equivalent
potentials.
An analytically solvable coupled-channel model has been employed in
order to implement the derivative expansion to higher orders
explicitly.
We have introduced a generalized derivative expansion to avoid a
problem which arises when the na\"ive derivative expansion is applied
to the present model due to its unsmooth NBS wave functions.
The convergence of the new expansion has been studied numerically.

We have observed that the generalized derivative expansion of the
non-local potentials converges quite well such that proper scattering
phase shift is extracted, although the functional structure of the
potentials is not uniquely determined.
In addition to the energy region where the agreement of the phase
shift is ensured by construction, the agreement extends to the higher
energies, suggesting that the extrapolation of the phase shift by the
generalized derivative expansion works successfully.
Moreover, we have observed that the convergence is improved by
properly choosing interpolating fields (the field admixture parameter
$q$) and/or the Gaussian expansion scale ($\rho$) in the generalized
derivative expansion.

\begin{acknowledgments}
We thank  Mr. K.~Hiranuma for  the discussions  at early stage  of the
study.
This work was supported by JSPS KAKENHI Grand Numbers JP25400244 and
JP25247036, and by MEXT as ``Priority Issue on Post-K computer''
(Elucidation of the Fundamental Laws and Evolution of the Universe)
and JICFuS.

\end{acknowledgments}

\appendix

\section{\label{sec:A-GDE}ADDITIONAL COMMENTS ON GENERALIZED DERIVATIVE EXPANSION}

\subsection{\label{sec:A-GDE-ms}Derivation of Generalized Derivative Expansion}
A general potential $V(x_1,x_2)\equiv \Braket{x_1|V|x_2}$ is
reparameterized in the coordinate space and in the momentum space as
\begin{align}
\label{eq:Vparameterization}
  \begin{split}
        {\mathcal V}(R,r) &\equiv \Braket{x_1|V|x_2}, \\
  \tilde{\mathcal V}(k,P) &\equiv \Braket{p_1|V|p_2},
  \end{split}
\end{align}
respectively, where the coordinates $r$ and $R$, and the momenta $k$
and $P$ are defined as
\begin{align}
  \begin{split}
    r = x_1 - x_2, & \hspace{10pt}
    R = x_1, \\
    k = p_1 - p_2, & \hspace{10pt}
    P = p_2.
  \end{split}
\end{align}
The two representations in Eqs.~\eqref{eq:Vparameterization} are
related to each other via the Fourier transformation
\begin{equation}
  \label{eq:FourierTransformation}
  {\mathcal V}(R,r)
  = \iint \frac{dk dP}{2\pi} e^{ikR}\, \tilde{\mathcal V}(k,P)\, e^{iPr}.
\end{equation}

First we derive the na\"ive derivative expansion \eqref{eq:NaiveDE}.
For   this   purpose,  we   replace   all   the  $P$   dependence   in
$\tilde{\mathcal V}(k,P)$  by  $\hat{P}=-i\partial  /  \partial  r$,  and  then
complete the integration over $P$ to have
\begin{equation}
  {\mathcal V}(R,r)
  =
  {\mathcal U}_0(R,-i\partial / \partial r)\, \delta(r),
\end{equation}
where
\begin{equation}
  {\mathcal U}_0(R, -i\partial / \partial r)
  \equiv
  \int dk\,
  e^{ikR}\,
  \tilde {\mathcal V}(k, -i\partial / \partial r).
\end{equation}
We  expand ${\mathcal  U}_0(R,-i\partial  / \partial  r)$  as a  power
series of the derivative $\partial/\partial r$.
To respect the time-reversal symmetry, ${\mathcal V}(R,r)$ have to be
real-valued.  Hence the power series is real at each order, and we
have
\begin{equation}
  {\mathcal U}_0(R,-i\partial / \partial r)
  =
  \sum_{n=0}^{\infty} u_{n}(R) \left(\frac{\partial}{\partial r} \right)^n,
\end{equation}
which leads to the (na\"ive) derivative expansion formula:
\begin{equation}
  V(x_1,x_2) =
  \sum_{n=0}^\infty u_n(x_1) 
  \left(\frac{\partial}{\partial x_1}\right)^n
  \delta(x_1-x_2).
\end{equation}

Derivation of the generalized derivative expansion starts from
factorizing the momentum-space representation $\tilde {\mathcal V}(k,
P)$ according to
\begin{equation}
  \tilde{\mathcal V}(k,P)
  \equiv 
  \tilde{\mathcal V}_\rho(k,P)\exp\left\{-\frac{1}{4}\rho^2P^2\right\},
\end{equation}
where an arbitrary real parameter  $\rho$ is introduced.
This time, the P dependence of the function $\tilde{\mathcal
  V}_\rho(k,P)$ is replaced by $\hat P = -i\partial/\partial r$, while
that of the Gaussian function is left unchanged. Completing the
integration of $P$ in Eq.~\eqref{eq:FourierTransformation}, we have
\begin{equation}
  {\mathcal V}(R,r)
  =
  {\mathcal U}_\rho(R,-i\partial /\partial r) \delta_{\rho}(r),
\end{equation}
where
\begin{equation}
  {\mathcal U}_\rho(R, -i\partial/\partial r)
  \equiv
  \int dk\,
  e^{ikR}
  \tilde {\mathcal V}_\rho(k, -i\partial/\partial r),
\end{equation}
and $\delta_\rho(r)$ is defined in Eq.~\eqref{eq:Gaussian-kernel}.
Again the time-reversal symmetry implies real-valued coefficients of
${\mathcal U}_\rho(R,-i\partial/\partial r)$ in the power-series
expansion of $\partial/\partial r$, leading to the generalized
derivative expansion
\begin{equation}
  V(x_1,x_2) = 
  \sum_{n=0}^\infty v_n^{(\rho)}(x_1)
  \left(\frac{\partial}{\partial x_1}\right)^n
  \delta_\rho (x_1-x_2).
\end{equation}

\subsection{Need for replacement $\partial/\partial x \to \partial/\partial (x^2)$}
\label{sec:need-for-the-replacement}
We  consider the  generalized derivative  expansion with  the conventional
derivatives $\partial/\partial x,\, \partial^2/\partial x^2, \cdots$.
The Schr\"odinger equations for $N+1$ energy levels are expressed in a
matrix form as
\begin{equation}
  {\bf u}(x)
  =
  M(x)
  {\bf v}(x),
  \label{eq:before-replacement}
\end{equation}
with
\begin{align}
  {\bf u}(x)
  &\equiv
  \left(
  \begin{array}{c}
    (E_0 - H_0) \Psi(x;E_0) \\
    (E_1 - H_0) \Psi(x;E_1) \\
    \vdots                  \\
    (E_N - H_0) \Psi(x;E_N)
  \end{array}
  \right),
  \nonumber \\
  M(x)
  &\equiv
  \left(
  \begin{array}{cccc}
    \Phi_\rho(x;E_0) & \partial_x \Phi_\rho(x;E_0) & \cdots & \partial_x^N \Phi_\rho(x;E_0) \\
    \Phi_\rho(x;E_1) & \partial_x \Phi_\rho(x;E_1) & \cdots & \partial_x^N \Phi_\rho(x;E_1) \\
    \vdots & \vdots & \ddots & \vdots \\
    \Phi_\rho(x;E_N) & \partial_x \Phi_\rho(x;E_N) & \cdots & \partial_x^N \Phi_\rho(x;E_N) \\
  \end{array}
  \right),
  \nonumber \\
  {\bf v}(x)
  &\equiv
  \left(
  \begin{array}{c}
    v_0(x) \\
    v_1(x) \\
    \vdots \\
    v_N(x)
  \end{array}
  \right),
\end{align}
where the summation over $n$ in the generalized derivative
expansion~\eqref{eq:GeneralizedDEx} is truncated at order $N$.
Equation~\eqref{eq:before-replacement}  is  solved   for  ${\bf  v}(x)$  by
inverting $M(x)$ point-by-point.

As far as the even-parity sector is considered, the smoothed wave
function $\Phi_\rho(x;E)$ behaves as
\begin{equation}
  \Phi_\rho(x;E)
  =
  \sum_{l=0}^{\infty} C_l(E) x^{2l}
\end{equation}
around $x=0$, and its $k$-th order derivative reads
\begin{equation}
  (\partial_x)^k\Psi_\rho(x;E)
  =
  \sum_{l=0}^{\infty} C_l(E) \cdot (\partial_x)^k x^{2l}.
\end{equation}
If $k$ is odd, $(\partial_x)^k \Psi_\rho(x;E)$ is of $\mathcal{O}(x)$,
so that the even-numbered columns of $M(x)$ vanish at $x=0$.
Hence the inversion of $M(x)$ fails and the coefficients ${\bf v}(x)$
diverges at $x=0$.
For  numerical  calculations,  such superficial divergence should be
removed explicitly.

This problem can be avoided when basis $\left(\partial/\partial
x\right)^k$ are replaced by $\left(\partial/\partial (x^2)
\right)^k$ ($k=0,1,\cdots$).
The two basis are related to each other by linear transformation
\begin{equation}
  \left(
  \frac{\partial}{\partial (x^2)}
  \right)^n
  =
  \sum_{m=0}^n
  \alpha_{nm}(x) \frac{\partial^m}{\partial x^m},
\end{equation}
where
\begin{align}
\begin{split}
  a_{nm}(x)
  &\equiv
  a_{nm}/x^{2n-m}
  \hspace{2em} \mbox{for  } n\ge 0, \\ 
  a_{0m}
  &\equiv
  \left\{
  \begin{array}{lcl}
    1 & \mbox{for} & m = 0 \\
    0 & \mbox{for} & m \neq 0
  \end{array}
  \right. , \\
  a_{1m}
  &\equiv
  \left\{
  \begin{array}{lcl}
    1/2 & \mbox{for} & m=1 \\
    0   & \mbox{for} & m\neq 1
  \end{array}
  \right. , \\
  a_{n+1,m}
  &\equiv
  \frac{a_{n,m-1}}{2}
  +
  \frac{m - 2n}{2} a_{n,m}
  \hspace{2em}
  \mbox{for  }
  n \ge 1.
\end{split}
\end{align}
Since $\alpha_{nm}(x)$ thus defined is a lower triangular matrix with
diagonal entries $\alpha_{nn}(x) = 1/(2x)^n$, $\alpha(x)$ is
invertible for $0 < |x| < \infty$.  Hence,
Eq.~\eqref{eq:before-replacement} is equivalent to
\begin{equation}
  {\bf u}(x) = \tilde M(x) \tilde {\bf v}(x),
  \label{eq:after-replacement}
\end{equation}
where
\begin{eqnarray}
  \tilde M(x)
  &\equiv&
  M(x) \alpha(x)^T
  ,\\
  \label{eq:vn-relation_xandx2}
  \tilde {\bf v}(x)
  &\equiv&
  \left(
  \begin{array}{c}
    \tilde v_0(x) \\
    \tilde v_1(x) \\
    \vdots        \\
    \tilde v_N(x)
  \end{array}
  \right)
  \equiv
  (\alpha(x)^T)^{-1} {\bf v}(x).
\end{eqnarray}
Notice   that  Eq.~\eqref{eq:after-replacement}   is  nothing   but  the
Schr\"odinger equations with the generalized derivative expansion where
the  conventional  derivatives  $\partial/\partial x$  are  replaced  by
$D_x\equiv \partial/\partial (x^2)$.
Although ${\bf v}(x)$ are singular at $x=0$, linear
combination~\eqref{eq:vn-relation_xandx2} cancels terms with negative
powers of $x$, leaving singularity-free coefficients $\tilde{\bf
  v}(x)$.
The replacement $\left(\partial/\partial x\right)^n \to
\left(\partial/\partial (x^2)\right)^n$ rearranges the coefficients
but does not change the non-local potential $V^{(N)}(x,x')$.

We take the $N=2$ case as an example. Matrix $\alpha(x)$ in this case
is given as
\begin{equation}
\alpha(x)=
\begin{pmatrix}
  1 & 0               & 0 \\
  0 &  \frac{1}{2x}   & 0 \\
  0 & -\frac{1}{4x^3} & \frac{1}{4x^2}
\end{pmatrix},
\end{equation}
which leads to 
\begin{align}
\begin{split}
\frac{\partial}{\partial (x^2)} 
  &=   \frac{1}{2x} \frac{\partial}{\partial x}, \\
\frac{\partial^2}{\partial (x^2)^2}
  &= - \frac{1}{4x^3} \frac{\partial}{\partial x} 
     + \frac{1}{4x^2} \frac{\partial^2}{\partial x^2}.
\end{split}
\end{align}
The relation between ${\bf v}(x)$ and $\tilde{\bf v}(x)$ then reads
\begin{align}
\begin{split}
  \tilde{v}_0(x) &= v_0(x), \\
  \tilde{v}_1(x) &= 2xv_1(x) + 2v_2(x), \\
  \tilde{v}_2(x) &= 4x^2 v_2(x),
\end{split}
\end{align}
so that negative powers of $x$ in ${\bf v}(x)$ are all cancelled in
$\tilde{\bf v}(x)$.

\section{\label{sec:A-GDE-TBC}Need for Twisted Boundary Condition}

We         consider         the         Schr\"odinger 
Eqs.~\eqref{eq:before-replacement}  with   the  generalized  derivative
expansion    of    the   conventional    derivatives    $\partial_x,
(\partial_x)^2, \cdots$, in the interval $-L<x<+L$.
Note that, in the even-parity sector, the following relation is
satisfied:
\begin{equation}
  \label{eq:parity-of-derivatives}
  [(\partial_x)^k\Phi_\rho](-x; E)
  =
  \left\{
  \begin{array}{lcl}
    + [(\partial_x)^k\Phi_\rho](x; E) & \mbox{for} & \mbox{even  } k
    \\
    - [(\partial_x)^k\Phi_\rho](x; E) & \mbox{for} & \mbox{odd   } k
  \end{array}
  \right.
\end{equation}
Now, suppose that the periodic boundary condition (PBC) is imposed on
the wave functions:
\begin{equation}
  \label{eq:PBCforphi}
  \Phi_\rho(x + 2L; E) = \Phi_\rho(x; E).
\end{equation}
Conditions~\eqref{eq:parity-of-derivatives} and \eqref{eq:PBCforphi}
imply
\begin{equation}
  (\partial_x)^k\Phi_\rho(x=L; E) = 0
  \hspace{2em}\mbox{for  odd } k
  \label{eq:periodic}
\end{equation}
at any energy.  Thus the even-numbered columns of the matrix $M(x)$
vanish at $x=\pm L$ so that $M(x)$ is not invertible.
A similar problem arises when the anti-periodic boundary condition
(APBC) is imposed such that
\begin{equation}
  \label{eq:APBCforphi}
  \Phi_\rho(x + 2L; E) = -\Phi_\rho(x; E).
\end{equation}
In this case the odd-numbered columns of $M(x)$ containing even-order
derivatives vanish, and $M(x)$ is uninvertible.

The above argument also holds when the conventional derivatives
$\left(\partial/\partial x\right)^n$ are replaced by $D_x^n$, since the
relation $\mbox{det}(\tilde M(x)) =
\mbox{det}(M(x))\mbox{det}(\alpha(x))$ implies that $\tilde M(x)$ is
uninvertible if $\mbox{det}M(x)=0$.

With large enough $L$, the potential is expected to become negligibly
small near the boundary.  However, as far as $L$ is finite,
non-invertible nature of $M(x)$ is still troublesome for numerical
determination of the potential.
Although the origin of the problem is similar to the one considered in
Appendix~\ref{sec:need-for-the-replacement}, we do not employ  similar
strategy,  because  it  results in  unwanted  $L$  dependence  in  the
potential.

An alternative solution is to impose the twisted boundary
conditions~\eqref{eq:TBC} with $\theta=\pi/2$. With this boundary
condition, the even-parity solution (the real part) is smoothly
connected with the odd-parity solution (the imaginary part) at the
boundary.  The derivatives of $\Phi_\rho(x;E)$ vanish only
accidentally and $E$-dependently, so that $M(x)$ is safely invertible
at $x=\pm L$.

\section{\label{sec:zero-determinant}Zero-Determinant Problem}

So far we have discussed the cases where the matrix inversion in 
Eq.~\eqref{eq:regular-part} is safely taken.
We have introduced the generalized derivative expansion since the
Birse model wave functions are not smooth at $x=\pm R$, i.e., at the
edge of the square-well potential.
The $x^2$-derivative operator $D= \partial/\partial (x^2)$ and the TBC
have been employed to avoid the problems at $x=0$ and $x=\pm L$,
respectively.
Each of these problems happens at a specific point in space and can be
explicitly avoided; however, in some cases an additional problem
arises at accidental points, which should be overcome one-by-one by
another technique.

The quantity of central importance is the determinant
\begin{equation}
\label{eq:detxq}
  \mbox{det}M(x;q)= \mbox{det}
  \begin{pmatrix}
    \Phi_q(x;E_0) & \cdots & D^N \Phi_q(x;E_0) \\
    \vdots & \ddots & \vdots \\
    \Phi_q(x;E_N) & \cdots & D^N \Phi_q(x;E_N)
  \end{pmatrix}.
\end{equation}
Let us suppose that the matrix $M$ is invertible in the whole spatial
region when we take $q=q_0$, i.e.
\begin{equation}
  \label{eq:h0def}
  h_0(x)\equiv \mbox{det}M(x;q_0) \neq 0
  \hspace{2em}\mbox{for  all  } x.
\end{equation}
We then try to improve the convergence of the generalized derivative
expansion by changing the field admixture parameter by $\delta q\equiv
q-q_0$, so that the NBS wave function reads
\begin{equation}
  \label{eq:wf_dev_from_q0}
  \Phi_q(x)
  =
  \Phi_{q_0}(x) + \delta q \phi_1(x),
\end{equation}
where $\phi_1(x)\equiv \int dx' \delta_\rho (x-x') \psi_1(x')$, the
Gaussian-smoothed function of $\psi_1(x)$, behaves in the qualitatively
same way as $\psi_1(x)$ itself, i.e., 
$\phi_1(x)\sim e^{-\gamma |x|}$.
By substituting Eq.~\eqref{eq:wf_dev_from_q0} in Eq.~\eqref{eq:detxq},
we find that the determinant is in general expressed as an (N+1)-th
order polynominal of $\delta q$:
\begin{equation}
  \label{eq:det-dqpows}
  \mbox{det}M(x;q)
  =
  h_0(x) + h_1(x)\delta q + \cdots + h_{N+1}(x) (\delta q)^{N+1}.
\end{equation}
The functions $h_n(x)$ ($n=0,\cdots,N+1$) are all real and continuous
in $x$, each of which involves n-fold product of $\phi_1(x)$ and/or
their derivatives.
Recalling that $\psi_1(x)$ (and thus $\phi_1(x)$) vanishes
exponentially at large distance, we find
\begin{equation}
  \label{eq:det-largex}
  \mbox{det}M(x;q) \simeq h_0(x),
\end{equation}
for large enough $x$, so that the matrix $M(x;q)$ is invertible.
On the other hand, Eq.~\eqref{eq:det-largex} cannot be satisfied for
small $x$, where the higher-order terms in Eq.~\eqref{eq:det-dqpows}
give non-vanishing contribution.
In principle, in some cases of $\delta q \neq 0$, there might be one
(or more) small $x$ where the determinant satisfies
$\mbox{det}M(x;q)=0$.
These zeros result in singular behavior in the coefficients of the
derivative expansion, $v_n(x)$, as we have discussed before.
The choice of $q$ value is restricted to a certain region to avoid
this problem, and the allowed region shall in general be smaller for
larger truncation order $N$, since the dependence in $q$ becomes
greater in Eq.~\eqref{eq:det-dqpows}.

The same argument is also valid in lattice QCD.
We consider constructing the HAL QCD potential for a system of two
hadrons, $H_A$ and $H_B$.
By using local interpolating fields $\hat{A}_0$ for $H_A$ and
$\hat{B}$ for $H_B$, we define the NBS wave function $\Psi_0({\bm x})$
as
\begin{equation}
  \Psi_0 ({\bm x})
  \equiv \Braket{0 | \hat{A}_0({\bm x}+{\bm y}) \hat{B}({\bm y}) | \Psi}.
\end{equation}
Meanwhile, we try to improve the convergence of the derivative
expansion by arranging the coupling of operator $\hat{A}_0$ to
the excited states of $H_A$ in the following way.
We first take a linear combination of $\hat{A}_0$ and another
interpolating field $\hat{A}_1$ for $H_A$ to construct an operator
$\hat{A}'$, which does not couple to the lowest-energy state of the
quantum numbers of $H_A$, $\ket{H_A;ground}$:
\begin{equation}
  \Braket{0 | \hat{A}' | H_A; ground} = 0.
\end{equation}
It is used to compose a new interpolating field
\begin{equation}
  \hat{A}_r \equiv \hat{A}_0 + r \hat{A}',
\end{equation}
with arbitrary parameter $r$.
We obtain the NBS wave function $\Psi_r({\bm x})$ with interpolating
fields $\hat{A}_r$ and $\hat{B}$,
\begin{equation}
  \Psi_r ({\bm x})
  \equiv \Braket{0 | \hat{A}_r({\bm x}+{\bm y}) \hat{B}({\bm y}) | \Psi},
\end{equation}
which asymptotically approaches $\Psi_0({\bm x})$, since the admixture
of $\hat{A}'$ is suppressed at large distance.
By comparing $\Psi_0({\bm x})$ and $\Psi_r({\bm x})$ to the NBS wave
functions in the previous argument with $q=q_0$ and $q=q_0+r$,
respectively, we find that the same problem can also occur in the
lattice QCD cases.

There is a way to overcome the problem in virtue of the introduction
of the generalized derivative expansion.
Recall the fact that the functional form of the non-local
potential varies depending on the expansion scale $\rho$.
Variance in $\rho$ can shift the zeros of $\mbox{det}M(x;\rho)$ (here
we explicitly indicate the $\rho$ dependence of $M$) as
\begin{align}
\begin{split}
  \mbox{det}M(x=a_1;\rho_1)=0 &\to
  \mbox{det}M(x=a_2;\rho_2)=0, \\
  a_1 \neq a_2,& \hspace{4pt} \rho_1 \neq \rho_2.
\end{split}
\label{eq:change_rho_zeros}
\end{align}
Moreover, if we determine more than two non-local potentials
$V^{(\rho)}(x,x')$ with different choices of $\rho$ from the same
$(N+1)$ NBS wave functions, all of those potentials manifestly
reproduce the input NBS wave functions.
It means that, $V^{(\rho)}(x,x')$ acts in the same way on $\Psi(x;E)$
for $E\leq E_N$, regardless of $\rho$.  Then it follows that a
weighted average of the potentials
\begin{align}
  \overline{V}(x,x') 
  &\equiv 
  \int d\rho\, w(x;\rho) V^{(\rho)}(x,x'), \\
  &\int d\rho\, w(x;\rho) = 1,
  \label{eq:weight_potential_average}
\end{align}
also reproduces the correct NBS wave functions for these energies.  
Note that the weight function $w(x;\rho)$ can be chosen totally
arbitrarily as far as the normalization
condition~\eqref{eq:weight_potential_average} is satisfied.

To be more specific, let us consider the case indicated in
Eq.~\eqref{eq:change_rho_zeros}, and assume that the corresponding
non-local potentials $V^{(\rho_1)}(x,x')$ and $V^{(\rho_2)}(x,x')$ are
obtained from the same NBS wave functions, despite being singular only
at $x=a_1$ and $x=a_2$, respectively.
We employ the weight function
\begin{equation}
w(x;\rho)
= \lambda(x)\delta(\rho-\rho_1)+(1-\lambda(x))\delta(\rho-\rho_2),
\end{equation}
so that
\begin{equation}
\overline{V}(x,x') = \lambda(x)V^{(\rho_1)}(x,x')+(1-\lambda(x))V^{(\rho_2)}(x,x').
\end{equation}
The function $\lambda(x)$ is taken to be smooth and to satisfy the
conditions $\lambda(x)=0$ for $x\simeq a_1$ and $\lambda(x)=1$ for
$x\simeq a_2$.
Then the singularities at $x=a_1$ and $x=a_2$ are both stamped out, so
that $\overline{V}(x,x')$ is singularity-free.
In actual calculations, the function $g(x)$ defined by
\begin{align}
\begin{split}
  g(x)
  &\equiv
  f(x-a)/(f(x-a)+f(b-x)), \\
  f(x)
  &\equiv 
  \begin{cases}
    \exp(-1/x) & \mbox{for  } x>0 \\
    0          & \mbox{for  } x\leq 0
  \end{cases},
\end{split}
\end{align}
will be useful to give $\lambda(x)$ a specific expression, which
satisfies $g(x\leq a)=0$ and $g(x\geq b)=1$, and is smoothly connected
in $a<x<b$.

By properly choosing the weight function $w(x;\rho)$, we can always
remove all the singularities from the HAL QCD potential, while
ensuring that it reproduces the correct NBS wave functions.
The above argument also implies that we need special care in
discussing the structure of a non-local potential, since a HAL QCD
potential can be an artificial patchwork of arbitrary pieces of
wave-function-equivalent potentials with different structures.

\section{\label{sec:A-BMS}Analytic Solutions}

\subsection{\label{sec:A-BMS-1}Solutions with TBC}
Coupled-channel Eqs.~\eqref{eq:CCeq} can be solved analytically in
a finite spatial interval with TBC~\eqref{eq:TBC}.
Here we provide the exact expression of the solution, since it is
rather complicated.


In energy region $0<E<\Delta$, the solution is given as:
\begin{align}
\psi_0(x) &= \begin{cases}
  (A\cos{\alpha x} + B\sin{\alpha |x|}) + C\sin{\alpha x}  & 0<|x|<R \\
  (D\cos{\beta x} + E\sin{\beta |x|}) \\ 
       \hspace{5mm} + (Fsgn(x)\cos{\beta x} + G\sin{\beta x}), & R<|x|<L
  \end{cases} \nonumber \\ \label{eq:BMsolution-a}
\psi_1(x) &= (H\cosh{\gamma x} + I\sinh{\gamma |x|}) + J\sinh\gamma x,
\end{align}
where $sgn(x)$ denotes the sign function
\begin{align}
  sgn(x) = \begin{cases}
    +1 & x>0 \\
    0  & x=0 \\
    -1 & x<0
    \end{cases}.
\end{align}

The associated momenta and the coefficients are given as follows.
\begin{equation}
\alpha \equiv \sqrt{M(E+V_0)},
\beta \equiv \sqrt{ME},
\gamma \equiv \sqrt{M(\Delta-E)},
\end{equation}
\begin{align}
\begin{split}
A &= (-iJ)\frac{\gamma}{Mg}(\cos\theta/\cosh{2\gamma L} -1), \allowdisplaybreaks\\
B &= (-iJ)\frac{Mg}{\alpha}\tanh{2 \gamma L}, \allowdisplaybreaks\\
C &= \frac J{\sin\theta}
           \frac{D_0\sinh{2\beta L} - E_0(\cos\theta + \cos{2\beta L})}
        {\sin{\alpha R}\sin{\beta R}+\frac{\alpha}{\beta}\cos{\alpha R}\cos{\beta R}}, 
        \allowdisplaybreaks\\
D &= (-iJ)D_0, \allowdisplaybreaks\\
E &= (-iJ)E_0, \allowdisplaybreaks\\
F &= \frac J{\sin\theta}\left(-D_0(\cos\theta-\cos{2\beta L})+E_0\sin{2\beta L}\right),
\allowdisplaybreaks\\
G &= \frac J{\sin\theta}\left(D_0\sin{2\beta L}-E_0(\cos\theta+\cos{2\beta L})\right),
\allowdisplaybreaks\\
H &= (-iJ)\tanh{2\gamma L}, \allowdisplaybreaks\\
I &= (-iJ)(\cos\theta/\cosh{2\gamma L} -1),
\end{split}
\end{align}
\begin{widetext}
\begin{align*}
D_0 &= \frac{\gamma}{Mg} \left(\cos{\alpha R}\cos{\beta R} 
       +\frac{\alpha}{\beta}\sin{\alpha R}\sin{\beta R}\right) 
        \left(\frac{\cos\theta}{\cosh{2\gamma L}} -1\right)
       +
        \frac{Mg}{\alpha}\left(\sin{\alpha R}\cos{\beta R}-\frac{\alpha}{\beta}
        \cos{\alpha R}\sin{\beta R}\right) \tanh{2\gamma L}, \\
E_0 &= \frac{\gamma}{Mg}\left(\cos{\alpha R}\sin{\beta R}
       -\frac{\alpha}{\beta}\sin{\alpha R}\cos{\beta R}\right)
        \left(\frac{\cos\theta}{\cosh{2\gamma L}}-1\right)
       +
        \frac{Mg}{\alpha}\left(\sin{\alpha R}\sin{\beta R}+\frac{\alpha}{\beta}
        \cos{\alpha R}\cos{\beta R}\right) \tanh{2\gamma L}.
\end{align*}
The allowed energy eigenstates satisfy the relation
\begin{equation}
\mbox{det}
\begin{pmatrix}
  +D_0(\cos\theta-\cos{2\beta L}) - E_0\sin{2\beta L} ,& 
  \sin{\alpha R}\cos{\beta R} - \frac{\alpha}{\beta}\cos{\alpha R}\sin{\beta R} \\
  -D_0\sin{2\beta L} + E_0(\cos\theta+\cos{2\beta L}) ,&
  \sin{\alpha R}\sin{\beta R} + \frac{\alpha}{\beta}\cos{\alpha R}\cos{\beta R}
\end{pmatrix}
= 0.
\label{eq:BMsolution-b}
\end{equation}
\end{widetext}
Be aware that, if we take the phase factor $J$ to be pure imaginary,
i.e. $J=i$, the parity-even and the parity-odd parts are separated
out.

We refrain from writing down the whole expressions for $-V_0<E<0$ and
$E<-V_0$ to avoid redundancy.
Instead, we note that the solution for $-V_0<E<0$ is obtained through
replacement
\begin{align}
\begin{split}
    \beta &\to \beta' \equiv i\sqrt{-ME}, \\
    \cos{\beta x} &\to \cosh{\beta' x}, \\
    \sin{\beta x} &\to i\sinh{\beta' x},
\end{split}
  \label{eq:momentum-replacement1}
\end{align}
in each relevant quantity in
Eqs. (\ref{eq:BMsolution-a}-\ref{eq:BMsolution-b}).
For  $E<-V_0$, further replacement
\begin{align}
\begin{split}
  \alpha &\to \alpha' \equiv i\sqrt{-M(E+V_0)}, \\
  \cos{\alpha x} &\to \cosh{\alpha' x}, \\
  \sin{\alpha x} &\to i\sinh{\alpha' x}
\end{split}
\end{align}
is necessary as well as
replacement~\eqref{eq:momentum-replacement1}.

\subsection{\label{sec:A-BMS-2}Scattering Phase Shift in the Infinite Volume}

It is less time-consuming to solve the coupled-channel
Eqs.~\eqref{eq:CCeq} in the infinite volume.
We give the exact expression of the scattering phase shift extracted
from this analytic solution as follows:
\begin{widetext}
\begin{equation}
\tan\delta_{exact}
= \frac{
      \frac{\gamma}{Mg}\left(\cos{\alpha R}\sin{\beta R}
                            -\frac{\alpha}{\beta}\sin{\alpha R}\cos{\beta R}\right)
     -\frac{Mg}{\alpha}\left(\sin{\alpha R}\sin{\beta R}
                            +\frac{\alpha}{\beta}\cos{\alpha R}\cos{\beta R}\right)
  }{
      \frac{\gamma}{Mg}\left(\cos{\alpha R}\cos{\beta R}
                             +\frac{\alpha}{\beta}\sin{\alpha R}\sin{\beta R}\right)
     -\frac{Mg}{\alpha}\left(\sin{\alpha R}\cos{\beta R}
                             -\frac{\alpha}{\beta}\cos{\alpha R}\sin{\beta R}\right)
  }.
\end{equation}
\end{widetext}

\end{document}